\documentclass[11pt,fleqn]{article}
\usepackage{amssymb}
\usepackage{amsfonts}
\usepackage{amsmath}
\usepackage{amsthm}
\usepackage{graphicx}
\usepackage{epsfig}
\usepackage{psfrag}
\usepackage{color}
\usepackage{dsfont}
\makeatletter
\xdef\@endgadget#1{{\unskip\nobreak\hfil\penalty50\hskip1em\hbox{}\nobreak
    \hfil#1\parfillskip=0pt\finalhyphendemerits=0\par}}

\usepackage{graphicx,color}
\usepackage{epstopdf}

\usepackage{cite}

\usepackage{subcaption}

\usepackage{amsfonts}

\usepackage{amsmath,amssymb}
\usepackage{amsthm}

\def\R{\mathbb{R}}

\newcommand{\be} {\begin{equation}}
\newcommand{\ee} {\end{equation}}

\newtheorem{theorem}{Theorem}
\newtheorem{prop}{Proposition}
\newtheorem{property}{Property}
\newtheorem{assumption}{Assumption}
\newtheorem{remark}{Remark}
\newtheorem{definition}{Definition}
\newtheorem{lemma}{Lemma}

\usepackage{adjustbox}

\def\BibTeX{{\rm B\kern-.05em{\sc i\kern-.025em b}\kern-.08em
    T\kern-.1667em\lower.7ex\hbox{E}\kern-.125emX}}

\title{\LARGE \bf Convergent Weight and Activation Dynamics in Memristor Neural Networks}

\author{Mauro Di Marco, Mauro Forti, Luca Pancioni,\\ Giacomo Innocenti, Alberto Tesi
\thanks{M. Di Marco, M. Forti and L. Pancioni are with the Department of Information Engineering and Mathematics, University of Siena, v. Roma 56 - 53100 Siena, Italy, e--mail: {\tt\small mauro.dimarco@unisi.it, mauro.forti@unisi.it, luca.pancioni@unisi.it}.
G. Innocenti and A. Tesi are with the Department of Information Engineering,
University of Florence, via S. Marta 3 - 50139 Firenze, Italy, e--mail: {\tt\small giacomo.innocenti@unifi.it,alberto.tesi@unifi.it}.
}
}

\begin{document}

\maketitle
\thispagestyle{empty}
\pagestyle{empty}

\begin{abstract}
Convergence of dynamic feedback neural networks (NNs), as the Cohen-Grossberg, Hopfield and cellular NNs, has been for a long time a workhorse of NN theory. Indeed, convergence in the presence of multiple stable equilibrium points (EPs) is crucial to implement content addressable memories and solve several other signal processing tasks in real time.
There are two typical ways to use a convergent NN, i.e.: a) let the activations evolve while maintaining fixed weights and inputs (activation dynamics) or b) adapt the weights while maintaining fixed activations (weight dynamics). As remarked in a seminal paper by Hirsch, there is another interesting possibility, i.e., let the neuron interconnection weights evolve while simultaneously running the activation dynamics (weight-activation dynamics). The weight-activation dynamics is of importance also because it is more plausible than the other two types for modeling neural systems.
The paper breaks new ground by analyzing for the first time in a systematic way the convergence properties of the weight-activation dynamics for a class of memristor feedback dynamic NNs. The main result is that, under suitable assumptions on the structure of the memristor interconnections, the solutions (weights and activations) converge to an EP, except at most for a set of initial conditions with zero measure. The result includes the most important case where the NN has multiple stable EPs.
\end{abstract}


\section{Introduction}
\label{sect:intro}

The importance of convergence of dynamic feedback neural networks (NNs), such as the classic Cohen-Grossberg \cite{CG83}, Hopfield \cite{Hop84} and cellular NNs \cite{CY88a}, cannot be overemphasized. It is indeed a needed property in order to implement content addressable memories (CAMs), in which case the NNs need to
possess multiple asymptotically stable equilibrium points (EPs) in correspondence with the memorized patterns. Convergent NNs also play a pivotal role in solving combinatorial optimization problems and
image processing, pattern recognition and classification tasks in real time. Convergence has been the workhorse of dynamic feedback NN theory for several years, see, e.g.,
\cite{Hay99,Hop84,Hir89,DFGP12a,CG83,CY88a,FT04,CR05,di2017memristor,cfc2020,
Ou20231518,Liu20231098,10144925}, and references therein. Recently, there has been a remarkable resurgence of interest in convergence, further boosted by the Nobel prize awarded to J. Hopfield for the expected role of the celebrated NN carrying his name in futuristic computers and AI \cite{Yuan202312949}.

Dynamic feedback NNs are characterized by two  main structural elements, i.e., the neuron activations corresponding for instance to voltages across capacitors and the feedback interconnection weights
implemented via conductances. There are two typical processing schemes implemented with a
convergent NN.

a) The most typical and widely investigated scheme is to let the activations evolve with time towards an equilibrium point (EP) while maintaining fixed weights, parameters and inputs, which is
referred to as \emph{activation dynamics}. All the fundamental results on convergence previously referenced to indeed pertain to convergent activation dynamics.

b) A second possibility is to adaptively change the weights while
maintaining fixed the activations with the goal to obtain a specific kind of dynamic behavior
(\emph{weight dynamics}). For instance, the goal may be to obtain a given set of stable equilibria and shape of their attraction basins by adapting the weights of a Hopfield NN \cite{pineda1987generalization,zhang2013common,scellier2017equilibrium,liao2018reviving,ganjidoost2024online}. Another example amounts to obtain desired image processing capabilities in a cellular NN by steering the trajectories so that they approximate a prescribed set of trajectories via adaptation of the NN interconnection cloning templates \cite{nossek1996design,adhikari2018building}.

As pointed out in the incipit of the introduction of the seminal review paper
by Hirsch \cite{Hir89}, there is also a third interesting little explored possibility, i.e., to
adapt the weights while simultaneously running the activation dynamics. This yields
a dynamical system in the Cartesian product of activations and weights, which
describes the \emph{weight-activation dynamics}. There are key motivations to investigate
the weight-activation dynamics. First of all, due to the interplay between activations and
weights, the weight-activation dynamics is expected to be richer and it may
include as special cases the other two types. Moreover, the weight-activation dynamics
is more plausible to model real neural systems than the other two types also from a biological viewpoint.
Indeed, it is well-known that in the brain the interconnection strengths invariably link
and adapt in correspondence to the neuron activities and conversely \cite{markram2012spike,van2017rewiring}.
To the authors' knowledge, although several years have passed since the
publication of the quoted paper by Hirsch, convergence of the weight-activation dynamics continues to be not much explored
so far and no general results are available in the literature.
One key reason is that it is intrinsically more difficult to study than convergence of the
activation or the weight dynamics. Furthermore, for structural reasons,
dynamic NNs as the Cohen-Grossberg, Hopfield and cellular NNs,
are characterized in their original form by fixed weights and it is
not obvious how to realistically replace them with weights that can
dynamically change.

Goal of this paper is to introduce a general class of NNs with
memristor interconnections (MNNs) where we allow both the activations and weights to vary with time and to
study some fundamental convergence properties of the weight-activation dynamics.
Memristor has been theoretically introduced by Chua in \cite{Chua1971} as the fourth fundamental passive circuit element in addition to the resistor, capacitor and inductor.
After the discovery of memristive behavior at nanoscale \cite{Williams2008}, there has been a widespread and worldwide burst of interest in the theory and applications of memristors in view of their use as memories that resemble the function of biological synapses and as basic building blocks of neuromorphic brain-like computers \cite{huang2021editorial,Sirakoulis2022717,SpecIssue-InMemory2023,Duan2024}. A memristor is a nonlinear resistor with a state dependent conductance (or memductance) which accounts for the whole history of the applied voltage. When the memristor is used as a synaptic connection in a NN, we can envisage two main modes of operation. In the first mode, the memristor is used as a programmable resistor, namely, the memductance is set at some desired value with the application of supra-threshold voltages and it is guaranteed that the activation voltages stay below threshold, so that the memductance remains fixed while the activations evolve with time (activation dynamics), see, e.g., \cite{yang2022ResProgMem}, and references therein. On the other hand, the case we are interested in this paper is that where during the activation dynamics the voltages may exceed the threshold, i.e., they are such that the memductances vary with time. Now, the memristor is a dynamic element and we are in the case of weight-activation dynamics.

The paper considers a class of NNs with an additive interconnecting structure where the neuron connections are implemented via memristors.
It is stressed that we allow both the neuron activations and the memductances to vary with time (weight-activation dynamics). Memristors are assumed to be passive while there are positive memristor interconnections between distinct neurons. The main result in the paper (Theorem\ \ref{th:conv MNN}) is that, under a suitable irreducibility assumption on the memristor interconnecting structure, the MNN is almost convergent, i.e., the solutions (activations and weights) converge to an EP except at most for a set of initial conditions with zero measure in the sense of Lebesgue. This result cannot exclude the presence of non-convergent (oscillatory or chaotic) solutions. Yet, in the practical applications almost convergence is as useful as convergence since there is probability zero to pick at random initial conditions originating non-convergent solutions. To the authors knowledge, Theorem\ \ref{th:conv MNN} is the first general result on convergence concerning
weight-activation dynamics. We also stress that the result naturally includes cases where the MNNs possess multiple asymptotically stable EPs, which is crucial for instance to implement CAMs.
The main results are verified and illustrated via the implementation of some simple low-order
CAMs and via the solution of some simple processing tasks.

The proof of Theorem\ \ref{th:conv MNN} is based on the use of two main techniques. First of all, it is shown that a MNN admits invariants of motion and the state space in the voltage-current domain (VCD), i.e., in the Cartesian product of weights and activations, is foliated in invariant manifolds where the MNN obeys a reduced-order dynamical system. Thereafter, the Flux-Charge Analysis Method (FCAM) developed in \cite{Corinto-Forti-I,cfc2020} is used to obtain the reduced-order system on each manifold in the flux-charge domain (FCD). It is shown that, under the considered assumptions, such a system turns out to be a cooperative and irreducible system of differential equations so that we can make use of the theory developed in a series of seminal papers by Hirsch
\cite{Hir85,Hir84,Hir89} to prove convergence results on each manifold (FCD) and then for the MNN (VCD). We stress that the reduction in the FCD via FCAM is crucial since in the VCD the MNN is not described by a cooperative system of differential equations.

\subsection{Preliminaries}

We will make use of the following definitions of convergence. For more details,
we refer the reader to \cite{Hir89}. Consider the system of ordinary differential equations (ODEs)
\begin{equation}\label{ODE}
    \dot x = F(x)
\end{equation}
where $F \in C^1(\R^n)$ and assume all its solutions are bounded and hence defined for
any $t \ge 0$. Let us denote by $E=\{ x \in \R^n:F(x)=0 \}$ the set of equilibrium points
(EPs) of (\ref{ODE}).

\begin{definition}
\label{def:conv}
System\ (\ref{ODE}) is said to be:

i) convergent if each solution tends to an EP as $t \to \infty$;

ii) quasi-convergent if each solution tends to the set of equilibria $E$ as $t \to \infty$;

iii) almost convergent if, for almost all (a.a.)\ initial conditions in $\R^n$, in the sense of the Lebesgue measure, the solution tends to an EP as $t \to \infty$;

iv) almost quasi-convergent if, for a.a.\ initial conditions in $\R^n$, the solution tends to $E$ as $t \to \infty$.
\end{definition}

\section{MNN Model}
\label{sect:MNN}
An ideal flux-controlled memristor \cite{Chua1971} is a two-terminal electric element defined by the constitutive
relation $q=\hat q(\varphi)$, where the charge $q(t)=\int_{-\infty}^t i(\sigma)d \sigma$ is the integral of the current $i$, the flux $\varphi(t)=\int_{-\infty}^t v(\sigma)d \sigma$ is the integral of the voltage $v$
and $\hat q$ is a nonlinear function. Assuming $\hat q$ is $C^1$, by differentiating in time, we have that the memristor satisfies the state-dependent Ohm's law
$$
                        \left\{
                           \begin{array}{ll}
                             i = \hat q'(\varphi)v \\
                             \dot \varphi = v.
                           \end{array}
                         \right.
$$
The term $\hat q'(\varphi)$, whose dimension is Ohm$^{-1}$, is named memductance and it represents a state-dependent conductance taking into account via $\varphi$ the whole history of the voltage applied to the memristor.

Any physically implemented memristor is \emph{passive} since it does not contain a source of energy. It is shown in \cite[Th.\ 1]{Chua1971} that passivity is equivalent to $\hat q$ being a monotone non-decreasing function. Actually, in this paper we assume the memristor satisfies the slightly more restrictive condition $\hat q'(\varphi)>0$ for any $\varphi$, i.e., the memristor is strictly passive \cite{Riaza2011607}. We also note that for any practical memristor the memductance varies between a minimum value
$G_\mathrm{off}>0$, corresponding to the fully insulating case, to a maximum value $G_\mathrm{on}<\infty$,
$G_\mathrm{on}>G_\mathrm{off}$, in the fully conductive state \cite{cfc2020}. Throughout the paper we thus suppose that any memristor satisfies the next condition.
\begin{assumption}
\label{assu:mem} The nonlinear characteristic $\hat q \in C^1(\R)$ is such that $\hat q(0)=0$
and
$
0<G_\mathrm{off} \le \hat q'(\varphi)\le G_\mathrm{on}<\infty
$
for any $\varphi$. Moreover,
$
\lim_{\varphi \to \infty}\hat q'(\varphi)= G_+,
\lim_{\varphi \to -\infty}\hat q'(\varphi)= G_-
$
with $G_\mathrm{off} \le G_+ \le G_\mathrm{on}$, $G_\mathrm{off} \le G_- \le G_\mathrm{on}$.
\end{assumption}

As it was discussed before, this assumption is quite natural and indeed it is
satisfied by relevant memristor models as the celebrated HP memristor \cite{Williams2008} (see Sect.\ \ref{sect:HP mem} for more details) or a combination of HP memristors \cite{di2017memristor}.

\begin{figure}[t]
  \centering
  \includegraphics[width=0.87 \linewidth]{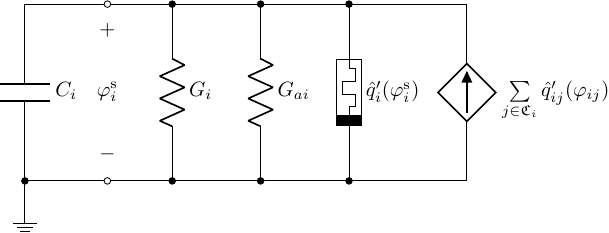}
\caption{\small Schematic of the $i$-th neuron of a MNN.}
\label{fig:MNN}
\end{figure}

Let us consider a NN given by an array of $n$ neurons. Figure\ \ref{fig:MNN} shows the schematic of the $i$-th neuron and the interconnections with the other neurons.
The neuron has a capacitor $C_i>0$, a passive conductance $G_i>0$, an active conductance $G_{ai}<0$ and a memristor $\hat q'_{i}$ representing the neuron self-connection. The interconnections of the $i$-th neuron with other neurons are represented by the currents $\hat q'_{ij}(\varphi_{ij})v_j$, $j \ne i$, injected via the voltage-controlled current-sources in Fig.\ \ref{fig:MNN}. Actually, these are obtained by interconnecting the $j$-th neuron, $j \ne i$, to the $i$-th neuron via memristors $\hat q'_{ij}$. Since under Assumption\ \ref{assu:mem} we have $\hat q'_{ij}(\varphi_{ij})>0$, we are dealing with \emph{positive interconnections between distinct neurons}. If there is no memristor connecting the $j$-th neuron with the $i$-th neuron, $j \ne i$, we simply let $\hat q'_{ij}=0$.
In overall, we have a NN having both memristor self-connections and interconnections between distinct neurons obtained via memristors. More precisely, it is seen that the considered architecture has a negative feedback obtained via $G_i$ and the passive memristor self-connections $\hat q'_{i}$ and also a positive feedback due to $G_{ai}$ and the memristor interconnections $\hat q'_{ij}$ with $j \ne i$.

While each neuron has a self connection, in the general case,
the MNN is not fully interconnected. We find it useful to introduce some notation to describe the interconnecting structure. Let $\mathfrak{N}=\{1,\dots,n\}$ and define $\mathfrak{C}_i=\{j \ne i:
q'_{ij}(\varphi_{ij}) \ne 0\}$, $i \in \mathfrak{N}$, which is the set of memristor interconnections with the $i$-th neuron. Also, let $n_{Ci}= {\rm card}\{ \mathfrak{C}_i\}$ and $n_C=\sum_i n_{Ci}$.
From Fig.\ \ref{fig:MNN} we have that in the standard voltage-current domain (VCD) the MNN satisfies
\begin{align}\label{MNN}
\begin{split}
C_i \dot v_i =& -(G_i+G_{ai})v_i-\hat q'_{i}(\varphi^\mathrm{s}_{i})v_i
 + \sum_{j \in \mathfrak{C}_i} \hat q'_{ij}(\varphi_{ij})v_j,i \in \mathfrak{N},\\
\dot \varphi^\mathrm{s}_{i}=& v_i, \ \ i \in \mathfrak{N},\\
\dot \varphi_{ij}=& v_j, \ \ i \in \mathfrak{N}, j \in \mathfrak{C}_i.
\end{split}
\end{align}
This is a system of $2n+n_C$ differential equations in the state variables given by the capacitor voltages $v_i$, $i \in \mathfrak{N}$, the self-connection memristor fluxes $\varphi^\mathrm{s}_{i}$, $i \in \mathfrak{N}$ and the interconnection memristor fluxes $\varphi_{ij}$, $i \in \mathfrak{N}, j \in \mathfrak{C}_i$.

It is worth to stress that in the MNN model we allow both the capacitor voltages and the memristor fluxes to vary with time.
According to the terminology in \cite{Hir89}, we will refer to the dynamics of $v_i$ as the \emph{activation dynamics} and to that of $\hat q'_i(\varphi^\mathrm{s}_i)$ and $\hat q'_{ij}(\varphi_{ij})$ as the \emph{weight dynamics}. The overall dynamics (\emph{weight-activation dynamics}) takes place in the product space $\R^n \times \R^{n} \times \R^{n_C}$. Clearly, we are dealing with the third case considered in the quoted paper by \cite{Hir89} (cf.\ Sect.\ \ref{sect:intro}).

The EPs of (\ref{MNN}) are such that $v_i=0$,
$\varphi^\mathrm{s}_i=\bar \varphi^\mathrm{s}_i$, $i \in \mathfrak{N}$ and $\varphi_{ij}=\bar \varphi_{ij}$,
$i \in \mathfrak{N}$, $j \in \mathfrak{C_i}$.  Note that there is a continuum of EPs and that
at each EP the capacitor voltages vanish while the memristor fluxes can assume any value.

\section{Convergent weight-activation dynamics of MNNs}
\label{sect:main}
In this section, we give the main results on convergence of the MNN. Their proof will be
postponed to Sect.\ \ref{sect:proof main}. First, we need a result ensuring
boundedness of solutions.

\begin{property}
\label{prope:bounded}
Suppose that the memristors satisfy Assumption\ \ref{assu:mem} and, moreover, assume
\begin{equation}\label{slope mem}
    \min \{ G_{i}^+,G_{i}^- \} > -(G_{ai}+G_i)+
\sum_{j \in \mathfrak{C}_i} \max \{ G_{ij}^+,G_{ij}^- \}
\end{equation}
$i \in \mathfrak{N}$. Then, any solution of (\ref{MNN}) is bounded and hence defined for $t \ge 0$.
\end{property}

\proof See Sect.\ \ref{sect:proof main}. \qed

To state the main result, we introduce the indicator matrix of the memristor interconnections.
Let $T=(T_{ij}) \in \R^{n \times n}$ be such that $T_{ii}=1$, $i \in \mathfrak{N}$, moreover,
 $T_{ij}=1$ if $\hat q'_{ij}(\varphi_{ij}) >0$, $i,j \in \mathfrak{\mathfrak{N}}$ (i.e.,
$i \in \mathfrak{N}$ and $j \in \mathfrak{C}_i$), whereas $T_{ij}=0$ otherwise. Note that due to Assumption\
\ref{assu:mem}, $T$ is a constant matrix which is independent of $\varphi_{ij}$.
Recall that matrix $T$ is said to be \emph{irreducible} if and
only if for every nonempty subset $I$ of $\mathfrak{N}$,
there exist indexes $i \in I$ and $j \in \mathfrak{N} \backslash I$ such that
$T_{ij} \ne 0$.

\begin{theorem}
\label{th:conv MNN}
Suppose the memristors satisfy Assumption\ \ref{assu:mem} and condition
(\ref{slope mem}) holds. If the indicator matrix $T$ is irreducible,
then the MNN (\ref{MNN}) is almost-convergent, i.e., each solution of (\ref{MNN}) converges to an EP
except at most for a set of initial
conditions in $\R^n \times \R^n \times \R^{n_C}$ with Lebesgue measure 0.
\end{theorem}

\proof See Sect.\ \ref{sect:proof main}. \qed

\begin{remark}
To the authors' knowledge, Theorem\ \ref{th:conv MNN} is the first general result on convergence for
NNs with interconnections implemented by means of ideal memristors where both the activation and interconnection weights are allowed to vary with time (weight-activation dynamics). A more detailed comparison with the literature
is provided in Sect.\ \ref{sect:disc}.
\end{remark}

\begin{remark}
Theorem\ \ref{th:conv MNN} means that, under the stated assumptions, the generic solution of (\ref{MNN})
is such that both the activations $v_i$, the fluxes $\varphi_i^\mathrm{s}$, $\varphi_{ij}$ and the weights, i.e., the memductances
$\hat q'(\varphi_i^\mathrm{s}),\hat q'(\varphi_{ij})$ converge to a constant asymptotic value. In particular, from Theorem\ \ref{th:conv MNN}, considering that any EP is such that the capacitor voltages $v_i=0$, $i \in \mathfrak{N}$, it follows that for any convergent solution $v_i$ vanish as $t \to +\infty$.
\end{remark}

\begin{remark}
Since the theorem guarantees almost convergence, we cannot exclude the presence of
nonconvergent solutions as periodic or even chaotic ones. However, they are not
stable and therefore they are not observable in practice. This means that
almost convergence is actually as useful as convergence in view of the application of the MNN to
signal processing tasks or to implement a CAM.
\end{remark}

\begin{remark}
We stress that the theorem includes the most relevant cases in view of the applications
where the MNN has multiple stable EPs, which is crucial for implementing
for instance CAMs (see the examples in Sect.\ \ref{sect:example}).
\end{remark}

\section{Proof of Main Convergence Result}
\label{sect:proof main}

At the core of the proof of Theorem\ \ref{th:conv MNN} there are two main techniques,
i.e., the use of FCAM \cite{Corinto-Forti-I,cfc2020} to obtain a dynamic description of the MNN in the FCD and then, the theory of cooperative and irreducible systems of ordinary
differential equations (ODEs) developed in a series of seminal papers by Hirsch \cite{Hir89}.

The proof is subdivided in a number of fundamental steps that can be summarized as follows:
1) Show that the MNN (\ref{MNN}) in the VCD admits invariants of motion and invariant manifolds (Sect.\ \ref{sect:inv}).
2) Use of FCAM to obtain the reduced-order system satisfied by the integral quantities
incremental flux and charge on each invariant manifold, i.e., in the FCD (Sect.\ \ref{sect:FCAM}).
3) Use the reduced-order system to prove Property\ \ref{prope:bounded}, i.e.,
boundedness of solutions of (\ref{MNN}) (Sect.\ \ref{sect:proof prope}).
4) Prove almost quasi-convergence of each reduced-order system on a given manifold
using the theory of cooperative and irreducible systems of ODEs (Sect.\ \ref{sect:almost quasi conv}).
5) Use Sard's lemma and Fubini's theorem to complete the
proof of Theorem\ \ref{th:conv MNN}, i.e., to show almost convergence
of (\ref{MNN}) (Sect.\ \ref{sect:proof th}).

\subsection{Invariants of Motion and Invariant Manifolds}
\label{sect:inv}

Let us define the vectors $v=(v_i)_{i \in \mathfrak{N}}$, the vector $\varphi^\mathrm{s}=(\varphi_i^\mathrm{s})_{i \in \mathfrak{N}} \in \R^n$ and $\varphi = ((\varphi_{1j})_{j \in \mathfrak{C}_1},\dots,
(\varphi_{nj})_{j \in \mathfrak{C}_n}) \in \R^{n_c}$.
Consider the system of differential equations (\ref{MNN}) describing the MNN in the VCD.
Since $\dot \varphi_i^\mathrm{s}=v_i$ and $\dot \varphi_{ij}=v_j$, we have from (\ref{MNN})
$$
C_i \dot v_i = -(G_i+G_{ai})\dot \varphi_i^\mathrm{s}-\frac{d}{dt} \hat q_{i}(\varphi_i^\mathrm{s}) +
\sum_{j \in \mathfrak{C}_i} \frac{d}{dt} \hat q_{ij}(\varphi_{ij}), \ \ i \in \mathfrak{N}.
$$
Therefore
$$
\frac{d}{dt}\left(C_i v_i +(G_i+G_{ai})\varphi_i^\mathrm{s}+ \hat q_{i}(\varphi_i^\mathrm{s}) -
\sum_{j \in \mathfrak{C}_i} \hat q_{ij}(\varphi_{ij})\right)=0
$$
for any $i \in \mathfrak{N}$. This implies that the functions
\begin{equation}\label{Qi}
    Q_i(v,\varphi^\mathrm{s},\varphi)=C_i v_i +(G_i+G_{ai})\varphi_i^\mathrm{s}+ \hat q_{i}(\varphi_i^\mathrm{s}) -
\sum_{j \in \mathfrak{C}_i} \hat q_{ij}(\varphi_{ij})
\end{equation}
where $i \in \mathfrak{N}$, are $n$ invariants of motion (or preserved quantities) for the MNN, i.e.,
they are constant along the solutions of the MNN.

Moreover, from the second and third equation in (\ref{MNN}),
$
\dot \varphi_j^\mathrm{s}=\dot \varphi_{ij}$, $i \in \mathfrak{N}, j \in \mathfrak{C_i}$.
Then, the functions
\begin{equation}\label{Qij}
    Q_{ij}(v,\varphi^\mathrm{s},\varphi)=\varphi_j^\mathrm{s}-\varphi_{ij},\ \ i \in \mathfrak{N}, j \in \mathfrak{C_i}
\end{equation}
are $n_C$ additional invariants of motion for the MNN.

Note that, in overall, the MNN admits $n+n_C$ invariants of motion.
Therefore, the state space of the MNN can decomposed (i.e.,
foliated) in $n+n_C$ invariant manifolds
\begin{align}
M(Q_0)=&\{ (v,\varphi^\mathrm{s},\varphi)\in \R^{2n+n_C}: Q_{i}(v,\varphi^\mathrm{s},\varphi)=Q_{i0}, i \in \mathfrak{N};
 \nonumber\\
       & Q_{ij}(v,\varphi^\mathrm{s},\varphi)=Q_{ij0}, i \in \mathfrak{N}, j \in \mathfrak{C_i}\}
\end{align}
where constants $Q_{i0}$ and $Q_{ij0}$ are given by
\begin{align}
\label{Q0}
Q_{i0}=& C_i v_{i0} +(G_i+G_{ai})\varphi^\mathrm{s}_{i0}+ \hat q_{i}(\varphi^\mathrm{s}_{i0})\\
 &-\sum_{j \in \mathfrak{C}_i} \hat q_{ij}(\varphi_{ij0}), \ \ i \in \mathfrak{N} \nonumber
\\
\label{Q0ij}
Q_{ij0}=& \varphi^\mathfrak{s}_{j0}-\varphi_{ij0},\ \ i \in \mathfrak{N}, j \in \mathfrak{C_i}
\end{align}
and they depend upon the initial conditions of the MNN
$v_{i0}=v_i(0)$, $i \in \mathfrak{N}$, $\varphi^\mathrm{s}_{i0}=\varphi^\mathrm{s}_{i}(0)$,
$i \in \mathfrak{N}$ and $\varphi_{ij0}=\varphi_{ij}(0)$, $i \in \mathfrak{N}$, $j \in \mathfrak{C}_i$.
Note that each manifold has dimension $n$ and it is uniquely
identified by the vector
$$
Q_0=((Q_{i0})_{i \in \mathfrak{N}}, (Q_{1j0})_{j \in \mathfrak{C}_1},\dots,
(Q_{nj0})_{j \in \mathfrak{C}_n}) \in \R^{n+n_c}
$$
which will be referred to henceforth as the manifold index.

For further reference,
we find it convenient to split $Q_0$ as
$Q_0=(Q_0^\mathrm{s},Q_0^\mathrm{i}) \in \R^{n+n_c}$,
where we have let
$$Q_0^\mathrm{s}=(Q_{i0})_{i \in \mathfrak{N}} \in \R^n$$
and
$$Q_0^\mathrm{i}=
(Q_{1j0})_{j \in \mathfrak{C}_1},\dots,
(Q_{nj0})_{j \in \mathfrak{C}_n}) \in \R^{n_c}.
$$

\subsection{Reduced-order System in the FCD via FCAM}
\label{sect:FCAM}

Since the state space of the MNN (\ref{MNN}) has dimension $2n+n_C$,
but the dynamics evolves on an $n$-dimensional invariant manifold $M(Q_0)$,
we expect that the dynamics on $M(Q_0)$ is described by
a reduced-order system, more precisely, an $n$-dimensional system.

\begin{figure}[t]
\centering
  \includegraphics[width=0.9\columnwidth]{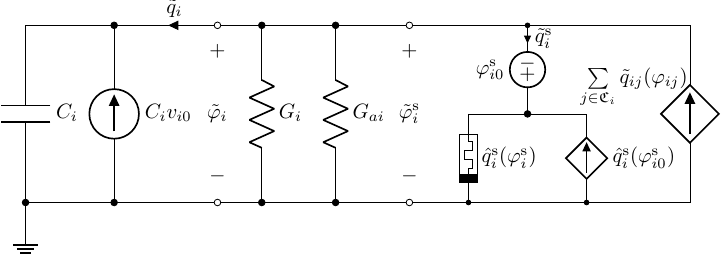}
\caption{\small Schematic of the $i$-th neuron of a MNN in the FCD as obtained via FCAM.}
\label{fig:MNNFC}
\end{figure}

FCAM is an effective technique to obtain the dynamic description on a manifold
via the use of the integral quantities given by the incremental flux
$\tilde \varphi(t)=\int_0^t v(\sigma)d\sigma$ and incremental charge
$\tilde q(t)=\int_0^t q(\sigma)d\sigma$ \cite{Corinto-Forti-I}. The schematic of a neuron
in the FCD is shown in Fig.\ \ref{fig:MNNFC}. According to FCAM, each element has
the incremental flux and charge at its terminals and it is described by a
relationship between these two quantities. In the figure, we used the equivalent
circuits in the FCD for the capacitor and memristors as given
in \cite[Sect.\ II]{Corinto-Forti-I}.
By using Kirchhoff-charge-law (K$q$L)
and Kirchhoff-flux-law (K$\varphi$L) in \cite{Corinto-Forti-I}, we obtain the following equations
\begin{align*}
    \tilde q_i + (G_i+G_{ai})\tilde \varphi_i+\tilde q^\mathrm{s}_{i}-
\sum_{j \in \mathfrak{C}_i} \tilde q_{ij}=0, \ \ i \in \mathfrak{N} \\
 \tilde \varphi_{ij}= \tilde \varphi_j = \tilde \varphi_j^\mathrm{s}, \ \ i \in \mathfrak{N}, j \in \mathfrak{C}_i\\
 \tilde q_i= C_i \frac{d}{dt} \tilde \varphi_i-C_i v_{i0}, \ \ i \in \mathfrak{N}\\
 \tilde q^\mathrm{s}_{i}= \hat q_{i}(\tilde \varphi^\mathrm{s}_{i}+\varphi^\mathrm{s}_{i0})-\hat q_i(\varphi^\mathrm{s}_{i0}), i \in \mathfrak{N}\\
 \tilde q_{ij}= \hat q_{ij}(\tilde \varphi_{ij}+\varphi_{ij0})-\hat q_{ij}(\varphi_{ij0}),
\ \ i \in \mathfrak{N}, j \in \mathfrak{C}_i.
\end{align*}
By substitution, considering that
$\varphi^\mathrm{s}_i=\tilde{\varphi}_i + \varphi^\mathrm{s}_{i0}$
and $\varphi_{ij}=\tilde{\varphi}_j + \varphi_{ij0}$, we obtain
\begin{align*}
\begin{split}
C_i \frac{d}{dt} \tilde \varphi_i=& -(G_i+G_{ai})\tilde \varphi_i-
\hat q_{i}(\tilde \varphi_i+\varphi^\mathrm{s}_{i0})\\
 &+\sum_{j \in \mathfrak{C}_i} \hat q_{ij}(\tilde \varphi_j+\varphi_{ij0})\\
 &+C_i v_{i0}+\hat q_{i}(\varphi^\mathrm{s}_{i0})
 -\sum_{j \in \mathfrak{C}_i} \hat q_{ij}(\varphi_{ij0})
\end{split}
\end{align*}
$i \in \mathfrak{N}$.
Finally, recalling that $\tilde \varphi_j^\mathrm{s}=\varphi_j^\mathrm{s}-\varphi_{j0}$,
$j \in \mathfrak{N}$, we have
\begin{align}
\label{MNNFC}
\begin{split}
C_i \dot \varphi_i^\mathrm{s}=f_i(\varphi^\mathrm{s};Q_0) =& -(G_i+G_{ai})\varphi_i^\mathrm{s}-\hat q_{i}(\varphi_i^\mathrm{s})\\
 &+\sum_{j \in \mathfrak{C}_i} \hat q_{ij}(\varphi_j^\mathrm{s}+\varphi_{ij0}-\varphi^\mathrm{s}_{j0})
 +Q_{i0}
\end{split}
\end{align}
where $i \in \mathfrak{N}$ and
$Q_{i0}, i \in \mathfrak{N}$, are defined in (\ref{Q0}).
This is a system of $n$ differential equations in the self-connection memristor fluxes $\varphi_i^\mathrm{s}$,
$i \in \mathfrak{N}$, describing the dynamics of the MNN on manifold $M(Q_0)$ in the FCD. The system is described by the vector field $f(\varphi^\mathrm{s};Q_0)=(f_i(\varphi^\mathrm{s};Q_0))_{i \in \mathfrak{N}}:\R^n \to \R^n$.
Note that in (\ref{MNNFC}) there are additive terms $Q_{i0}$, $i \in \mathfrak{N}$.
The EPs of (\ref{MNNFC}) are the vectors $\bar \varphi^\mathrm{s}\ \in \R^n$ satisfying $f(\bar \varphi^\mathrm{s};Q_0)=0$.

To obtain the relation between solutions of (\ref{MNNFC}) in the FCD and those of (\ref{MNN})
in the VCD, define the auxiliary function
$$
\Gamma(\varphi^\mathrm{s};Q_0)=
(v,\varphi^\mathrm{s},\varphi):\R^n \to \R^{2n+n_c}
$$
where
\begin{equation}\label{v0}
    v_{i}=\frac{1}{C_i}[Q_{i0} -(G_i+G_{ai})\varphi^\mathrm{s}_{i}- \hat q_{i}(\varphi^\mathrm{s}_{i}) +
\sum_{j \in \mathfrak{C}_i} \hat q_{ij}(\varphi_{ij})],\ \ i \in \mathfrak{N}
\end{equation}
and
\begin{equation}\label{Phi0}
    \varphi_{ij}=\varphi^\mathrm{s}_{j}-Q_{ij0},\ \ i \in \mathfrak{N}, j \in \mathfrak{C}_i.
\end{equation}
Given $Q_0 \in \R^{n+n_C}$, denote by $\varphi^\mathrm{s}(t;\varphi_0)$ the solution of (\ref{MNNFC}) with
initial conditions $\varphi^\mathrm{s}_0=(\varphi^\mathrm{s}_{i0})_{i \in \mathfrak{N}} \in \R^n$. Then,
exploiting the expression of the invariants of motion
(\ref{Qi}) and (\ref{Qij}), it is possible to prove that
$
(v(t),\varphi^\mathrm{s}(t),\varphi(t))=\Gamma(\varphi^\mathrm{s}(t);Q_0)
$
is the solution of (\ref{MNN}) with initial conditions
$\Gamma(\varphi^\mathrm{s}_0;Q_0)$.

\subsection{Proof of Property\ \ref{prope:bounded}}
\label{sect:proof prope}

First, we prove that the solutions of the reduced-order system~(\ref{MNNFC}) are bounded.
Without loss of generality, suppose $C_i=1$, $i \in \mathfrak{N}$. Let us consider function $V(\varphi^\mathrm{s})= \|\varphi^\mathrm{s}\|_\infty=\max_{k \in \{1,\dots, n\}}|\varphi^\mathrm{s}_i|: \R^n \to \R$, which is convex in $\R^n$ and radially unbounded. Let $\varphi^\mathrm{s}(t)$, $t\in [0,\tau]$, $0<\tau<+\infty$ be a solution of~(\ref{MNNFC}). Since $V(\cdot)$ is an absolutely continuous function on $[0,\tau]$, it is differentiable for a.a.\ $t\in[0,\tau]$. Let us set $\bar k=\arg\{\max_{k \in \{1,\dots, n\}}|\varphi^\mathrm{s}_i|\}$, which is (generally) time-dependent. It can be shown (see
Appendix\ A) that
\begin{align}
\begin{split}
\label{DVDT}
\dot V(\varphi^\mathrm{s}(t))  \le & -(G_{a \bar{k}}+G_{\bar{k}})|\varphi^\mathrm{s}_{\bar{k}}(t)|\\
 &- \min	\{ | \hat{q}_{\bar k} (\varphi^\mathrm{s}_{\bar k}(t)|, | \hat{q}_{\bar k} (-\varphi^\mathrm{s}_{\bar k}(t)| \} \\
 &+ \Sigma_{j \in \mathfrak{C}_{\bar k}} \max \{| \hat{q}_{\bar{k} j} (\varphi^\mathrm{s}_{\bar k}(t)-\varphi^\mathrm{s}_{j0} + \varphi_{\bar{k} j0}|,\\
 &| \hat{q}_{\bar{k} j} (-\varphi^\mathrm{s}_{\bar k}(t)-\varphi^\mathrm{s}_{j 0} + \varphi_{\bar{k} j0}| \} + |Q_{\bar{k} 0}|
\end{split}
\end{align}
for a.a.\ $t \in [0,\tau]$.
From~(\ref{slope mem}) there exists $\varepsilon>0$ such that, for any $\bar{k}$
\begin{equation}
\label{eq:epsilongain}
-\min \{ G_{\bar k}^+, G_{\bar k}^- \} -(G_{a \bar k}+G_{\bar k}) +
\sum_{j \in \mathfrak{C}_{\bar k}} \max \{ G_{\bar k j}^+,G_{\bar k j}^- \} < -\epsilon.
\end{equation}
Let $\lambda=\varepsilon/(n+1)$. Assumption\ \ref{assu:mem} implies that there exists $\varphi^\mathrm{s}_M(\varepsilon)>0$ such that, if $|\varphi^\mathrm{s}_{\bar k}|>\varphi^\mathrm{s}_M(\varepsilon)$, the inequalities
\begin{align}
\label{eq:Gselfini}
\begin{split}
|\hat{q}( \varphi^\mathrm{s}_{\bar k} )- G_{\bar k}^+\varphi^\mathrm{s}_{ \bar k } |< \lambda \varphi^\mathrm{s}_{ \bar k} & \ \mathrm{if} \ \varphi^\mathrm{s}_{ \bar k} >0 \\
|\hat{q}( \varphi^\mathrm{s}_{\bar k} )- G_{\bar k}^-\varphi^\mathrm{s}_{ \bar k } |< \lambda |\varphi^\mathrm{s}_{ \bar k}| & \ \mathrm{if} \ \varphi^\mathrm{s}_{ \bar k} <0
\end{split}
\end{align}
and
\begin{align}
\label{eq:Gini}
\begin{split}
|\hat{q}_{\bar{k} j}( \varphi^\mathrm{s}_{\bar k} -\varphi^\mathrm{s}_{j 0} + \varphi_{\bar{k} j 0})- G_{\bar k}^+\varphi^\mathrm{s}_{\bar k} |< \lambda \varphi^\mathrm{s}_{ \bar k} & \ \mathrm{if} \ \varphi^\mathrm{s}_{ \bar k} >0 \\
|\hat{q}_{\bar{k} j}( \varphi^\mathrm{s}_{\bar k} -\varphi^\mathrm{s}_{j0} + \varphi_{\bar{k}j0})- G_{\bar k}^- \varphi^\mathrm{s}_{ \bar k } |< \lambda |\varphi^\mathrm{s}_{ \bar k}| & \ \mathrm{if} \ \varphi^\mathrm{s}_{ \bar k} <0
\end{split}
\end{align}
hold. If we choose
$
\varphi_M(\varepsilon)= \max \left \{ \varphi^\mathrm{s}_M(\varepsilon), (n+1) |Q_{ \bar{k} 0}|/\varepsilon \right \},
$
then for any $\varphi^\mathrm{s}_{\bar k}(t)$ such that $|\varphi_{\bar k}^\mathrm{s}(t)|> \varphi_M(\varepsilon) $, using~(\ref{eq:Gselfini}) and~(\ref{eq:Gini}), we have
\begin{align*}
\dot V(\varphi^\mathrm{s}(t)) & \le  |\varphi_{\bar k}^\mathrm{s}(t)| \left ( -(G_{a \bar k} + G_{\bar{k}}) - \min \{ G_{i}^+ , G_{i}^- \} + \frac{\varepsilon}{n+1}  \right. \\
& \left. +\sum_{j \in \mathfrak{C}_i} \max \{ G_{ij}^+,G_{ij}^- \} + \frac{ n_{Ci} \varepsilon }{n+1} +\frac{\varepsilon}{n+1}  \right ).
\end{align*}
Using~(\ref{eq:epsilongain}), this boils down to
\begin{align*}
\dot V(\varphi^\mathrm{s}(t))& < \varepsilon |\varphi_{\bar k}^\mathrm{s}(t)|  \left(\frac{n_{Ci}+2}{n+1}-1\right) .
\end{align*}
Since $n_{Ci} \le n-1$ and $|\varphi^\mathrm{s}_{\bar k}(t)|= \| \varphi^\mathrm{s}(t) \|_\infty$, if $\| \varphi^\mathrm{s}(t)\|_\infty > \varphi_M(\varepsilon)$, then for a.a.\ $t\in [0,\tau]$ we have $\dot V(\varphi^\mathrm{s}(t)) < 0$. As a consequence, the set $B_{\varphi_M}= \{\varphi^\mathrm{s} \in \R^n: |\varphi^\mathrm{s}|\le \ \varphi_M(\varepsilon) \}$ is positively invariant and globally attractive for the solutions of~(\ref{MNNFC}). Therefore, these are bounded and defined for $t \ge 0$.

Consider now the solutions of (\ref{MNN}).
Recalling that $v_i = \dot{\varphi}^\mathrm{s}_i$, $\varphi_{ij}=\varphi^\mathrm{s}_j-\varphi^\mathrm{s}_{j0}+\varphi_{ij0}$ and considering that $\hat{q} \in C^1$, boundedness of $\varphi^\mathrm{s}$ in~(\ref{MNNFC}) implies also boundedness of $v_i$ and $\varphi_{ij}$ in~(\ref{MNN}). This completes the proof. \qed

\subsection{Almost Quasi-convergence on a Given Manifold}
\label{sect:almost quasi conv}

Consider system (\ref{MNNFC}) describing the
dynamics of the MNN on an invariant manifold in the VCD.
The following holds.

\begin{prop}
\label{prop:almost quasi conv Q0}
Suppose that the assumptions of Theorem\ \ref{th:conv MNN} are satisfied.
Then, for any $Q_0 \in \R^n$, system (\ref{MNNFC}) is almost quasi-convergent.
Moreover, for a.a.\ $Q_0 \in \R^{n+n_C}$, in the sense of the Lebesgue measure,
(\ref{MNNFC}) has isolated EPs and is almost convergent.
\end{prop}

\proof
From (\ref{MNNFC}), considering Assumption\ \ref{assu:mem}, we have
$$
\frac{\partial f_i(\varphi^\mathrm{s};Q_0)}{\partial \varphi_j^\mathrm{s}}=\hat q'_{ij}(\varphi_i^\mathrm{s}) \ge 0
$$
for any $\varphi^\mathrm{s} \in \R^n$ and any $i,j \in \mathfrak{N}$ with $i \ne j$, hence,
(\ref{MNNFC}) is a cooperative system of differential equations \cite{Hir89}.
Moreover, since the indicator matrix $T$ of the interconnections is irreducible,
it follows that the $n \times n$ Jacobian matrix
$
\left( \partial f_i(\varphi^\mathrm{s};Q_0)/\partial \varphi_j^\mathrm{s} \right)_{i,j \in \mathfrak{N}}
$
is irreducible for any $\varphi^\mathrm{s} \in \R^n$.
Therefore, (\ref{MNNFC}) is a cooperative and irreducible system of differential equations with
bounded solutions (Property\ \ref{prope:bounded}). Due to a classic result by Hirsch \cite{Hir85}, it follows that
for any $Q_0 \in \R^{n+n_c}$, system (\ref{MNNFC}) is almost quasi-convergent.

To complete the proof of the proposition we will exploit the
following auxiliary result.

\begin{lemma}[Sard]
\label{lemma:Sard}
Suppose that $z \in \R^k$ is a regular value of $F(\cdot;\cdot): \R^{m+p} \to \R^k$,
i.e., we have ${\rm rank}J_F(x,y)=k$ for any $(x,y) \in \R^{m+p}$ such that
$F(x,y)=z$. Then, for
a.a.\ $\alpha \in \R^p$, in the sense of the Lebesgue measure,
$z$ is a regular value of $F(\cdot,\alpha):\R^k \to \R^k$ .
\end{lemma}

Without loss of generality, let $C_i=1$, $i \in \mathfrak{N}$
and consider function $f(\cdot;\cdot):\R^{n+n_c} \to \R^n$, whose Jacobian is given by
$$
J_f(\varphi^\mathrm{s};Q_0)=\left (\frac{\partial f(\varphi^\mathrm{s};Q_0)}{\partial \varphi^\mathrm{s}}, \frac{\partial f(\varphi^\mathrm{s};Q_0)}{\partial Q_0}\right ).
$$
From (\ref{MNNFC}), $\partial f_i(\varphi^\mathrm{s};Q_0)/\partial Q_{i0}=1$, $i \in \mathfrak{N}$, hence there exists an $n \times n$ diagonal submatrix of $J_f(\varphi^\mathrm{s};Q_0)$ which is non-singular. Therefore, ${\rm rank}J_f(\varphi^\mathrm{s};Q_0)=n$ and any $z \in \R^n$ is a regular value of $f(\cdot;\cdot):\R^n \times \R^{n+n_c} \to \R^n$.
By applying Sard's lemma with $z=0$, we conclude that $z =0$ is a regular value of
$f(\cdot;Q_0):\R^n \to \R^n$ for a.a.\ $Q_0 \in \R^{n+n_c}$. This means that, for a.a.\ $Q_0 \in \R^{n+n_c}$, we have ${\rm rank}J_f(\bar \varphi^\mathrm{s};Q_0)=n$ at any $\varphi^\mathrm{s}$ such that $f(\bar \varphi^\mathrm{s};Q_0)=0$, i.e., at any EP of (\ref{MNNFC}). Therefore,
by the inverse function theorem any EP of (\ref{MNNFC}) is isolated. To conclude the proof of
Proposition\ \ref{prop:almost quasi conv Q0}, it suffices to recall that for isolated
equilibria quasi-convergence is equivalent to convergence \cite{Hir89}.
\qed

\subsection{Proof of Theorem\ \ref{th:conv MNN}}
\label{sect:proof th}
We are now ready to complete the proof of the main convergence result in
Theorem\ \ref{th:conv MNN}. Once more suppose, without loss of generality, that $C_i=1$,
$i \in \mathfrak{N}$.

Given any $Q_0 \in \R^{n+n_c}$, let $V(Q_0)$ be the set of initial conditions
$\varphi_0^\mathrm{s} \in \R^n$ for which the solution of (\ref{MNNFC}) does not converge to an EP. Due to
Proposition\ \ref{prop:almost quasi conv Q0}, we have $\mu_n(V(Q_0))=0$ for
a.a.\ $Q_0 \in \R^{n+n_c}$, where $\mu_n$ denotes the Lebesgue measure in $\R^n$.
If $\varphi_0^\mathrm{s} \in \R^n \backslash V(Q_0)$, then the corresponding solution  of (\ref{MNNFC}) converges to an EP, while it is not convergent otherwise.
Denote by $W$ the set of initial conditions in $\R^{2n+n_c}$ originating
solutions of (\ref{MNN}) that do not converge to an EP. We have
$$
W= \cup_{Q_0 \in \R^{n+n_c}} \{ (v_0,\varphi^\mathrm{s}_0,\varphi_0): \varphi^\mathrm{s}_0 \in V(Q_0) \}.
$$
To complete the proof let us show that $\mu_{2n+n_c}(W)=0$.
We have $\mu_{2n+n_c}(W)= \int_{\R^{2n+n_c}} \chi(x,y,z) dx dy dz$,
where the indicator function of $W$ is defined as $\chi(x,y,z)=1$ if $(x,y,z) \in W$
and $\chi(x,y,z)=0$ otherwise.

Consider the change of variables
$$
(x,y,z)= \Gamma(\varphi_0^\mathrm{s}; Q_0)= \Gamma(\varphi_0^\mathrm{s}; Q_0^\mathrm{s},Q_0^\mathrm{i})=(v_0,\varphi^\mathrm{s}_0,\varphi_0)
$$
where $v_0$ and $\varphi_0$ are defined in (\ref{v0}) and (\ref{Phi0}), respectively, whose Jacobian is
given by
\begin{align*}
 J_c=\left(
      \begin{array}{ccc}
      \frac{\partial v_0}{\partial \varphi^\mathrm{s}_0}   & \frac{\partial v_0}{\partial Q_0^\mathrm{s}}  & \frac{\partial v_0}{\partial Q_0^\mathrm{i}} \\
 \frac{\partial \varphi^\mathrm{s}_0}{\partial \varphi^\mathrm{s}_0}   & \frac{\partial \varphi^\mathrm{s}_0}{\partial Q_0^\mathrm{s}}  & \frac{\partial \varphi^\mathrm{s}_0}{\partial Q_0^\mathrm{i}} \\
\frac{\partial \varphi_0}{\partial \varphi^\mathrm{s}_0}   & \frac{\partial \varphi_0}{\partial Q_0^\mathrm{s}}  & \frac{\partial \varphi_0}{\partial Q_0^\mathrm{i}} \\
      \end{array}
    \right)
= \left(
      \begin{array}{ccc}
        \frac{\partial v_0}{\partial \varphi^\mathrm{s}_0} & E_n & \frac{\partial v_0}{\partial Q_0^\mathrm{i}} \\
         E_n  & 0 & 0 \\
         \frac{\partial \varphi_0}{\partial \varphi^\mathrm{s}_0}    & 0  & -E_{n_c}  \\
      \end{array}
    \right)
\end{align*}
where $E_n$ (resp., $E_{n_c}$) is the $n\times n$ (resp., $n_c \times n_c$ identity matrix).
It can be checked that $|\det J_c|=1$,
hence by applying the change of variables and using Fubini's theorem we obtain
\begin{align*}
\int_{\R^{2n+n_c}} \chi(x,y,z) dx dy dz
=&\int_{\R^{2n+n_c}} \chi(\Gamma(\varphi_0^\mathrm{s}))dv_0 d\varphi^\mathrm{s}_0 d\varphi_0\\
=&\int_{\R^{n+n_c}}
( \int_{\R^n} \chi(\Gamma(\varphi_0^\mathrm{s})) d\varphi_0^\mathrm{s}) dv_0 d\varphi_0.
\end{align*}
Since $\chi(\Gamma(\varphi_0^\mathrm{s})) =1 $ if and only if $\varphi_0^\mathrm{s} \in V(Q_0)$,
we have
$$
\int_{\R^n} \chi(\Gamma(\varphi_0^\mathrm{s})) d\varphi_0^\mathrm{s}
= \int_{\R^n} \chi(v_0,\varphi^\mathrm{s}_0,\varphi_0) d\varphi_0
=\mu_n (V(Q_0))=0.
$$
Hence, we proved that $\mu_{2n+n_c}(W)=0$. \qed

\section{Applications and Simulation Results}
\label{sect:example}
This section is divided into three main parts. First, we discuss some typical memristors models
that satisfy Assumption\ \ref{assu:mem} (Sect.\ \ref{sect:mem model}). Then, we consider some prototypical low-order MNNs and discuss the structure of stable and unstable EPs and the convergence properties towards stable EPs (Sect.\ \ref{sect:simple MNN}). Finally, to show the potential usefulness of the considered architectures, we consider an higher-order 2D MNN and its application in solving an elementary image processing task (Sect.\ \ref{sect:MNN image proc}).

\subsection{Some Typical Memristor Models}
\label{sect:mem model}

\subsubsection{Itoh-Chua Memristor}

\begin{figure}[t]
\begin{center}
\begin{subfigure}{0.49\columnwidth}
  % include first image
  \includegraphics[width=\linewidth]{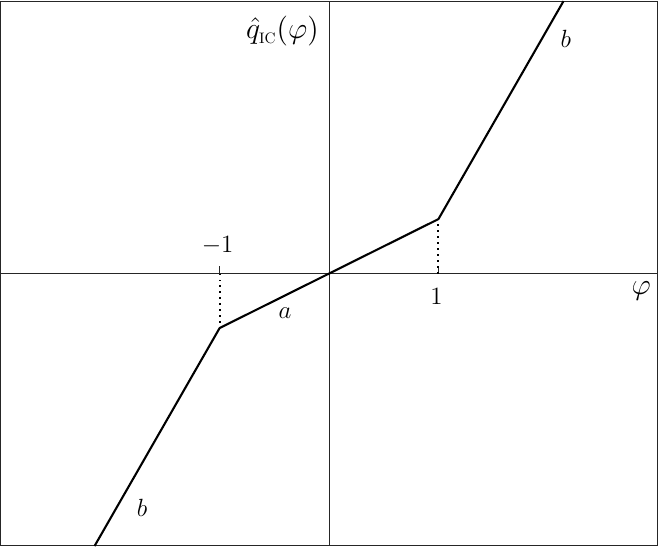}
  \caption{\small }
\end{subfigure}
\begin{subfigure}{0.49\columnwidth}
    % include second image
  \includegraphics[width=\linewidth]{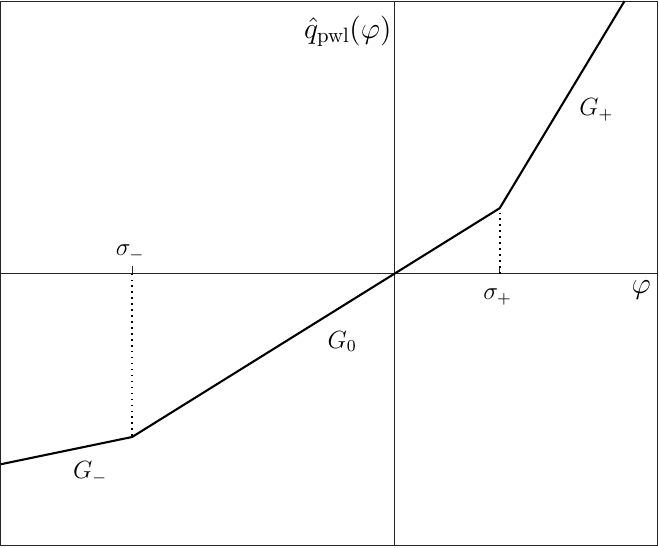}
  \caption{\small }
\end{subfigure} \\
\caption{\small (a) Three-segment pwl memristor characteristic in \cite{Itoh2008}. (b) General pwl memristor characteristic having three segments with positive slopes $G_-$,
$G_0$ and $G_+$ and corner points $\sigma_-$, $\sigma_+$.}
\label{fig:pwl mem}
\end{center}
\end{figure}

In the seminal paper \cite{Itoh2008}, a flux-controlled memristor has been considered which is modeled by the three-segment piecewise-linear (pwl) characteristic
$$
q= \hat q_\mathrm{IC}(\varphi)=b\varphi +\frac{1}{2}(a-b)(|\varphi+1|-|\varphi-1|)
$$
with $b>a >0$ (Fig.\ \ref{fig:pwl mem}(a)). Note that the memductance $\hat q'_\mathrm{IC}(\varphi)=b$ when $|\varphi|>1$, while
$\hat q'_\mathrm{IC}(\varphi)=a$ when $|\varphi|<1$. Henceforth, we will approximate $\hat q_\mathrm{IC}$ with a $C^1$ function obtained by smoothing the pwl characteristic in a small neighborhood around the corner points $\varphi_M=\pm 1$ so that we obtain a strongly passive memristor satisfying Assumption\ \ref{assu:mem} with
$G_\mathrm{off}=a$, $G_\mathrm{on}=b$ and $G_+=G_-=b$.

As it is done in the literature, we can also consider a more general class of three-segment pwl
memristor characteristics as shown in Fig.\ \ref{fig:pwl mem}(b). Assuming $\sigma_-<0$, $\sigma_+>0$, by using the canonical representation given in \cite{ChuaKang1977} we have
\begin{equation}\label{pwl mem}
 q=\hat q_\mathrm{pwl}(\varphi)=a+b\varphi+c_-|\varphi-\sigma_-|+c_+|\varphi-\sigma_+|
\end{equation}
where $b=(G_+ +G_-)/2$, $c_+=(G_+-G_0)/2$, $c_-=(G_0 -G_-)/2$ and
$a=[G_0(\sigma_+ +\sigma_-)-G_+ \sigma_+ -G_- \sigma_-]/2$.

\subsubsection{HP Memristor}
\label{sect:HP mem}
Let us consider the celebrated HP memristor proposed in \cite{Williams2008} as a model of a real metal-insulator-metal (MIM) nanostructure
\begin{align}\label{HP}
    v=& [R_\mathrm{on}x+R_\mathrm{off}(1-x)]i\\
    \dot x=& \beta W(x,p) i
\end{align}
where $x \in [0,1]$ is the normalized length of the conducting layer,
$0<R_\mathrm{on}<R_\mathrm{off}$ are the resistances in the fully conductive state ($x=1$) and
fully insulating state
($x=0$), respectively, and $\beta$ is a physical parameter.
Moreover, $W(\cdot,p)$ is a parametric window satisfying $W(0,p)=W(1,p)=0$ used
to guarantee that the state $x$ evolves in the range $[0,1]$. Henceforth, we will use
the classic window in \cite{Joglekar2009661} with $p=1$, i.e.
$
W(x,p)=1-(2x-1)^2.
$

The HP memristor is both charge- and flux-controlled. Its
flux-charge characteristic
$
q= \hat q_\mathrm{HP}(\varphi)
$
has the following properties \cite{DFT-NN-2017}.
We have $\hat q_\mathrm{HP} \in C^\infty$, $\hat q_\mathrm{HP}(0)=0$, moreover
$1/R_\mathrm{off} \le \hat q'_\mathrm{HP}(\varphi) \le 1/R_\mathrm{on}$, $\hat q''_\mathrm{HP}(\varphi)>0$
and $\lim_{\varphi \to \infty} \hat q_\mathrm{HP}(\varphi)=1/R_\mathrm{on}$,
$\lim_{\varphi \to -\infty} \hat q_\mathrm{HP}(\varphi)=1/R_\mathrm{off}$. Then, the HP
memristor satisfies Assumption\ \ref{assu:mem} with
$G_\mathrm{on}=1/R_\mathrm{on}$, $G_\mathrm{off}=1/R_\mathrm{off}$ and
$G_+=1/R_\mathrm{on}$, $G_-=1/R_\mathrm{off}$.

\subsection{A Simple MNN}
\label{sect:simple MNN}

Consider a MNN with $n=4$ neurons and suppose the memristor self-connections and interconnections are modeled by a three-segment pwl characteristic as in
(\ref{pwl mem}). First, we consider the dynamics of the MNN in the FCD, i.e., on a given manifold
(cf.\ Sect.\ \ref{sect:inv}), which is described by (\ref{MNNFC}).
Clearly, the pwl characteristics yield a partition of the state space $\R^4$ of (\ref{MNNFC})
on a manifold (cf.\ Sect.\ \ref{sect:inv}) in subregions where the MNN obeys a linear system.
Choose $C_i=1$, $i=1,\dots,4$.
For the memristor self-connections we let $G_-^\mathrm{s}=3$,
$G_0^\mathrm{s}=0.1$ and $G_+^\mathrm{s}=2.6$, whereas for the memristor interconnections,
$G_-=0.3$, $G_0=0.1$ and $G_+=0.4$. For all memristors, $\sigma_-=-1$ and $\sigma_+=1$.
The pwl characteristics are smoothed in a small interval around the corners $\pm 1$ in order that they satisfy Assumption\ \ref{assu:mem}. Moreover, we suppose $G_i+G_{ai}=G$, $i=1,\dots,4$.
The interconnections are defined via the indicator matrix (cf.\ Sect.\ \ref{sect:MNN})
$$
T=\begin{pmatrix}
    1 & 1 & 0 & 1 \\
    1 & 1 & 1 & 0 \\
    0 & 1 & 1 & 0 \\
    0 & 1 & 0 & 1 \\
  \end{pmatrix}.
$$
Since $T^{n-1}=T^3$ has all positive entries, $T$ is irreducible
\cite[Th.\ 6.2.23, p.\ 402]{horn2012matrix}.

The MNN (\ref{MNNFC}) is defined by the interconnecting structure $T$ and it depends upon the several parameters, namely, the coefficient $G$ of the linear part, the coefficients of the memristor interconnections, the manifold index $Q_0$ and the initial conditions for the memristors $\varphi_{i0}^\mathrm{s}$ and $\varphi_{ij0}$. Next, we numerically analyze some basic properties of the structure of the MNN EPs by varying some of these parameters.

\emph{Case\ 1)} Suppose $Q_0=0$, i.e., we consider the MNN dynamics on manifold $M(0)$. In order that the MNN has bounded solutions, from Property\ \ref{prope:bounded}, we need to have $G>-1.8$. Table~\ref{table:case1} provides the number $\Pi$ of partitions of the state space $\R^4$ of
(\ref{MNNFC}) due to the pwl memristor characteristics, the number $n_\mathrm{AS}$ of asymptotically stable EPs and the number
$n_\mathrm{u}$ of unstable EPs of the MNN, as a function of $G$, as obtained by means of a MATLAB program. Note that, while $\Pi$ remains constant, $n_\mathrm{as}$ and $n_\mathrm{u}$ largely vary with $G$.

\begin{table}
\begin{center}
$$
\begin{array}{|c|c|c|c|}
  \hline
  \mathbf{G} & \mathbf{\Pi} & \mathbf{n_\mathrm{as}} & \mathbf{n_\mathrm{u}} \\
  \hline
  -0.21 & 81 & 2 & 19 \\
  -0.22 & 81 & 5 & 24 \\
  -0.24 & 81 & 8 & 31  \\
  -0.25 & 81 & 9 & 34  \\
  -0.35 & 81 & 12 & 55  \\
  -0.4 & 81 & 15 & 62  \\
  -0.6 & 81 & 16 & 65  \\
  \hline
\end{array}
$$
\end{center}
\caption{\small Number of partitions of the state space $\R^4$ ($\Pi$),  number  of asymptotically stable EPs ($n_\mathrm{AS}$) and number of unstable EPs ($n_\mathrm{u}$) of the MNN, as a function of $G$ (Case 1).}
\label{table:case1}
\end{table}

\emph{Case\ 2)} As a second experiment,
we suppose that $G=-1$, while we vary the manifold index. More specifically,
we vary $Q^\mathrm{s}_0$ while taking $Q^\mathrm{i}_0=0$. For any assigned
$\varphi_{i0}^\mathrm{s}$, $i=1,\dots,4$, this can be done by choosing
$\varphi_{ij0}=\varphi_{j0}^\mathrm{s}$, $i=1,\dots,4$, $j \in \mathfrak{C}_i$
(cf.\ (\ref{Q0ij})) and finding $v_{i0}$, $i=1,\dots,4$, from (\ref{Q0}).
Let
$
Q^\mathrm{s}_0= \alpha(1, -1, 1, -1)^\top
$
and vary $\alpha$ in the interval $[0,3]$.
Table~\ref{table:case2} provides $\Pi$, $n_\mathrm{as}$ and $n_\mathrm{u}$
as a function of $Q_0$. While $\Pi$ is not affected by $Q_0^\mathrm{s}$, changing the
manifold index instead induces drastic changes in the number of stable and unstable EPs.

\begin{table}
\begin{center}
$$
\begin{array}{|c|c|c|c|}
  \hline
  \mathbf{\alpha} & \mathbf{\Pi} & \mathbf{n_\mathrm{as}} & \mathbf{n_\mathrm{u}} \\
  \hline
  0 & 81 & 16 & 65 \\
  0.3 & 81 & 15 & 64 \\
  0.6 & 81 & 11 & 50  \\
  0.7 & 81 & 7 & 30  \\
  1.2 & 81 & 6 & 9  \\
  1.4 & 81 & 3 & 2  \\
  3 & 81 & 1 & 0 \\
  \hline
\end{array}
$$
\end{center}
\caption{\small Number of partitions of the state space $\R^4$ ($\Pi$),  number  of asymptotically stable EPs ($n_\mathrm{AS}$) and number of unstable EPs ($n_\mathrm{u}$) of the MNN, as a function of $Q^\mathrm{s}_0$ (Case 2).}
\label{table:case2}
\end{table}

\emph{Case\ 3)}
In the third case, we suppose that $G=-1$ and vary $Q_0$ as follows. We let $Q_0^\mathrm{s}=0$,
while
$
Q_0^\mathrm{i}=\alpha(Q_{120},Q_{140},Q_{210},Q_{230},Q_{320},Q_{420})^\top
=\alpha(1,-0.5,0.6,-0.3,0.7,-0.4)^\top$
where $\alpha \in [0,1.6]$. Given $\varphi_{i0}^\mathrm{s}$, $i=1,\dots,4$, this can be done by choosing
$\varphi_{ij0}=\varphi_{j0}^\mathrm{s}-Q_{ij0}$, $i=1,\dots,4$, $j \in \mathfrak{C}_i$
(cf.\ (\ref{Q0ij})) and finding $v_{i0}$, $i=1,\dots,4$, from (\ref{Q0}).
The initial conditions for the memristor fluxes can be written as
\begin{align*}
 &\begin{pmatrix}
  \varphi^\mathrm{s}_{10} & \varphi_{120} & 0 & \varphi_{130} \\
  \varphi_{210} & \varphi^\mathrm{s}_{20} & \varphi_{230} & 0 \\
  0 & \varphi_{320} & \varphi^\mathrm{s}_{30} & 0 \\
  0 & \varphi_{420} & 0 & \varphi^\mathrm{s}_{40} \\
\end{pmatrix} \\
 &=\begin{pmatrix}
  \varphi^\mathrm{s}_{10} & \varphi^\mathrm{s}_{20}-\alpha & 0 & \varphi^\mathrm{s}_{40}+0.5 \alpha \\
  \varphi^\mathrm{s}_{10}-0.6 \alpha & \varphi^\mathrm{s}_{20} & \varphi^\mathrm{s}_{30}+0.3\alpha & 0 \\
  0 & \varphi^\mathrm{s}_{20}-0.7\alpha & \varphi^\mathrm{s}_{30} & 0 \\
  0 & \varphi^\mathrm{s}_{20}+0.4\alpha & 0 & \varphi^\mathrm{s}_{40}
\end{pmatrix}.
\end{align*}
From Table~\ref{table:case3} it is seen that varying $\alpha$ yields
drastic changes, not only in $n_\mathrm{as}$ and $n_u$, but also in the number $\Pi$ of subregions.

\begin{table}
\begin{center}
$$
\begin{array}{|c|c|c|c|}
  \hline
  \mathbf{\alpha} & \mathbf{\Pi} & \mathbf{n_\mathrm{as}} & \mathbf{n_\mathrm{u}} \\
  \hline
  0 & 81 & 16 & 65 \\
  2 & 1000 & 15 & 64 \\
  3 & 1125 & 10 & 55  \\
  4 & 900 & 7 & 32  \\
  5 & 1000 & 6 & 19  \\
  8 & 1125 & 4 & 3  \\
  10 & 1125 & 2 & 1 \\
  \hline
\end{array}
$$
\end{center}
\caption{\small Number of partitions of the state space $\R^4$ ($\Pi$),  number  of asymptotically stable EPs ($n_\mathrm{AS}$) and number of unstable EPs ($n_\mathrm{u}$) of the MNN, as a function of $Q_0^\mathrm{i}$ (Case 3).}
\label{table:case3}
\end{table}

\begin{figure}
\begin{center}
\begin{subfigure}{0.47\columnwidth}
  % include first image
  \includegraphics[width=\linewidth]{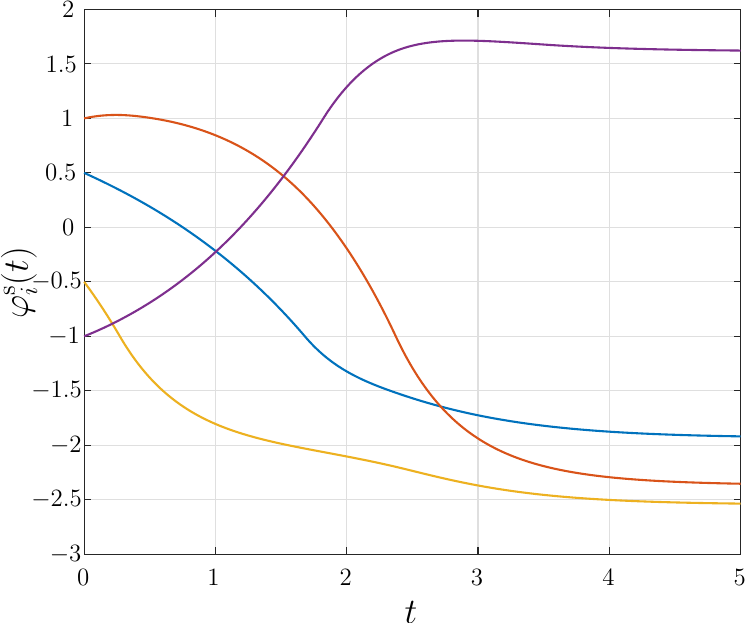}
  \caption{\small }
\end{subfigure}
\begin{subfigure}{0.47\columnwidth}
    % include second image
  \includegraphics[width=\linewidth]{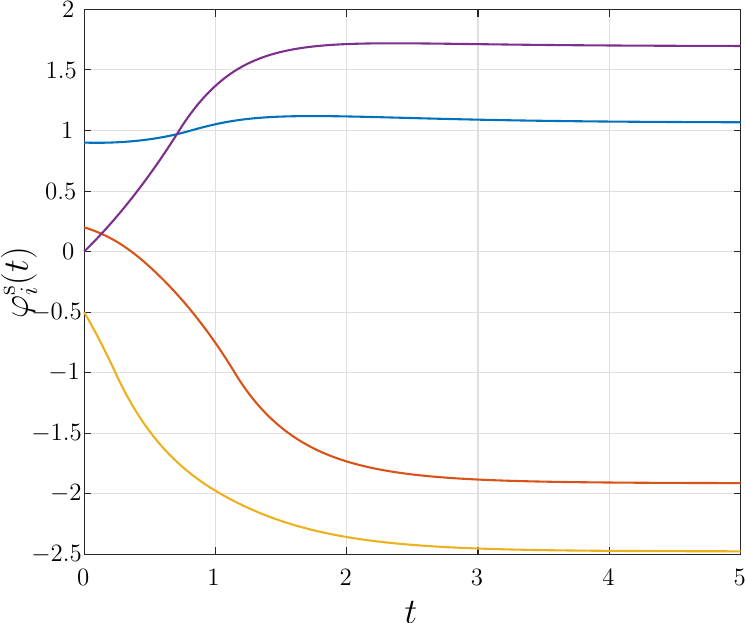}
  \caption{\small }
\end{subfigure} \\
\begin{subfigure}{0.47\columnwidth}
    % include first image
  \includegraphics[width=\linewidth]{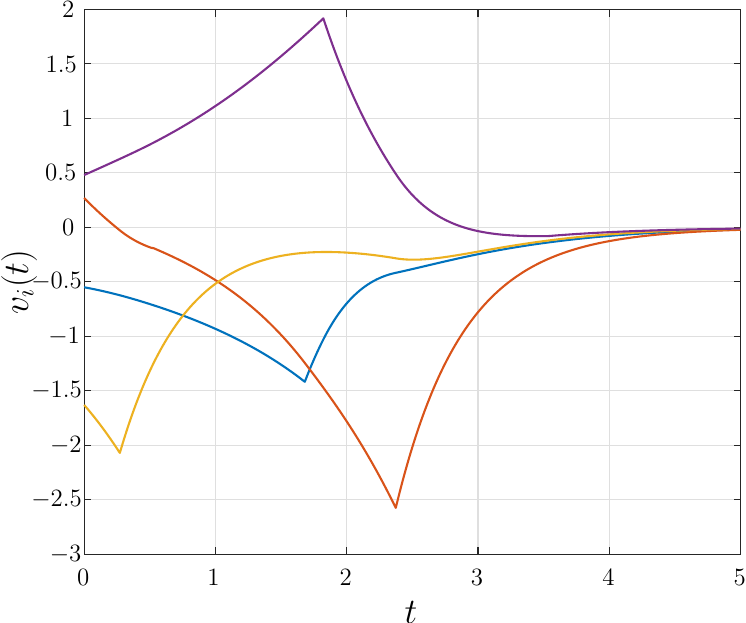}
  \caption{\small }
\end{subfigure}
\begin{subfigure}{0.47\columnwidth}
    % include second image
  \includegraphics[width=\linewidth]{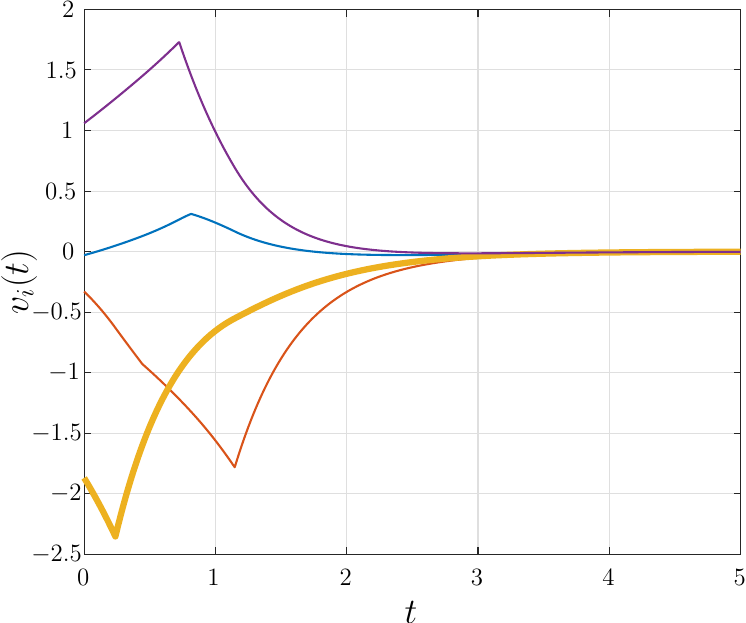}
  \caption{\small }
\end{subfigure}\\
\begin{subfigure}{0.47\columnwidth}
  % include first image
  \includegraphics[width=\linewidth]{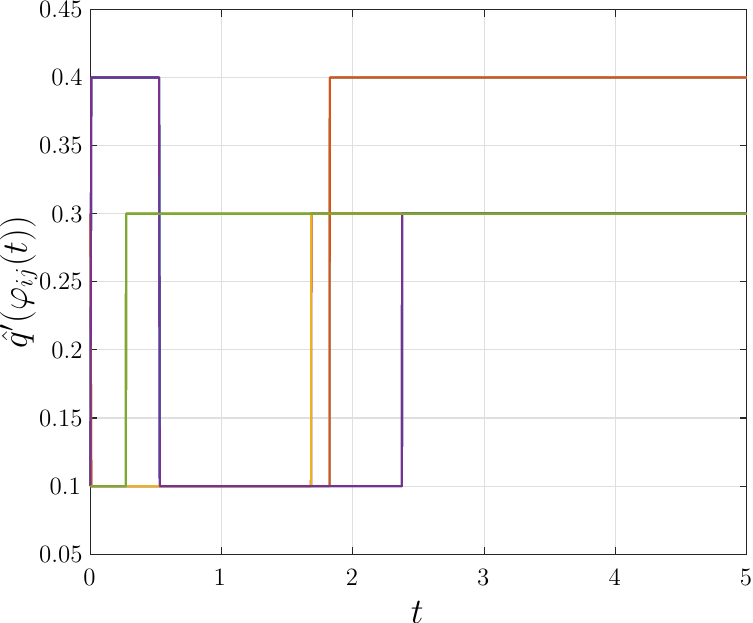}
  \caption{\small }
\end{subfigure}
\begin{subfigure}{0.47\columnwidth}
    % include second image
  \includegraphics[width=\linewidth]{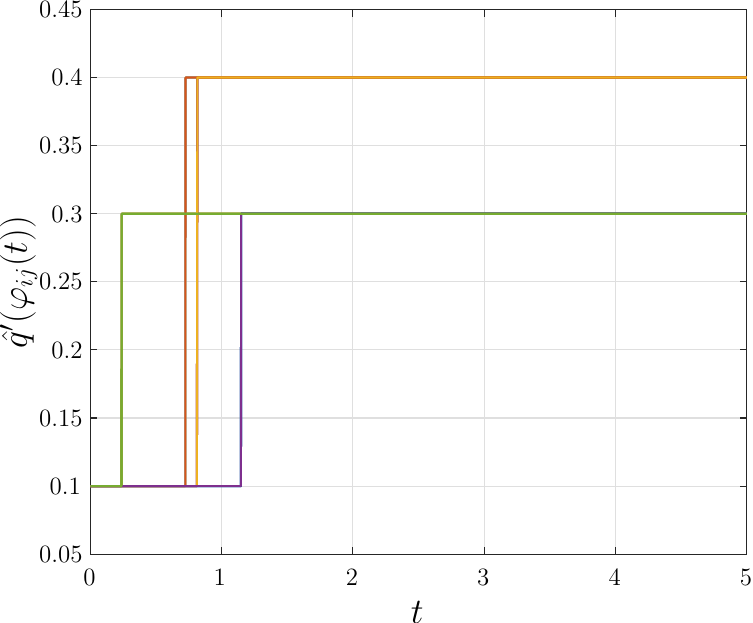}
  \caption{\small }
\end{subfigure} \\
\caption{\small (a), (b) Time-domain behavior of fluxes $\varphi_i^\mathrm{s}$ for a fourth-order MNN
with three-segments pwl memristor characteristics and two different sets of
initial conditions $\varphi_{i0}^\mathrm{s}$, $i=1,\dots,4$ ($\alpha=8$). (c), (d) Corresponding behavior of activations $v_i$ and (e), (f) of memductances $\hat q'(\varphi_{ij})$.}
\label{fig:trans beta1}
\end{center}
\end{figure}

We simulated with MATLAB the MNN (\ref{MNNFC}) in the FCD in Case\ 3).
Figures\ \ref{fig:trans beta1}(a), (b) show the time-domain
behavior of the neuron fluxes $\varphi_i^\mathrm{s}$ when $\alpha = 8$ for
two different initial conditions $\varphi_{i0}^\mathrm{s}$, $i=1,\dots,4$.
In each case we observe
convergence to an EP of (\ref{MNNFC}), in accordance with Proposition\ \ref{prop:almost quasi conv Q0}.
Let us now consider the corresponding MNN (\ref{MNN}) in the VCD and illustrate
its weight-activation dynamics in the space $\R^n \times \R^n \times \R^{n_c}=\R^4 \times
\R^4 \times \R^6$.
Figures\ \ref{fig:trans beta1}(c), (d) show the behavior of the neuron activations of (\ref{MNN}),
i.e., the memristor voltages $v_i$.
Note that $v_i$ tend to 0 as $t \to \infty$, in accordance with Theorem\ \ref{th:conv MNN}.
Finally, Figs.\ \ref{fig:trans beta1}(e), (f) show the behavior of the weights, i.e., the interconnection memductances
$\hat q'(\varphi_{ij})$. It is worth to observe that these are
subject to a number of switches between values $0.1, 0.3$ and $0.4$ before convergence.

\subsection{A Simple MNN with HP Memristors}

Here we consider a MNN as in case 3) of Sect.\ \ref{sect:simple MNN}, the main difference
being that the memristor interconnections are implemented with an HP memristor as in
Sect.\ \ref{sect:HP mem}. In particular, for the pwl memristor self-connections we
let $G_-^\mathrm{s}=8$, $G_0^\mathrm{s}=0.1$ and $G_+^\mathrm{s}=8$, moreover, $\sigma_-=-1$, $\sigma_+=1$.
For the HP memristor interconnections we choose $R_\mathrm{on}=0.2$, $R_\mathrm{off}=1$ and $\beta=0.1$.
Finally, we let $G=-1$.

\begin{figure}
\begin{center}
\begin{subfigure}{0.47\columnwidth}
  % include first image
  \includegraphics[width=\linewidth]{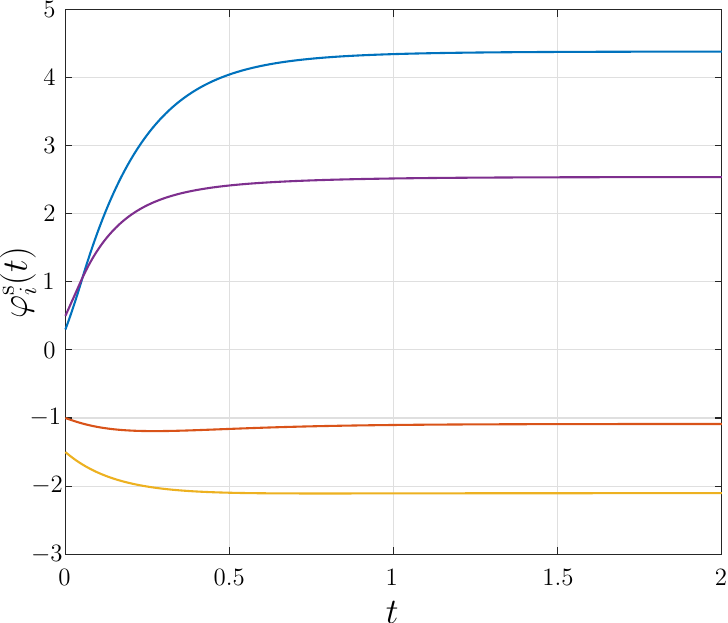}
  \caption{\small }
\end{subfigure}
\begin{subfigure}{0.47\columnwidth}
    % include second image
  \includegraphics[width=\linewidth]{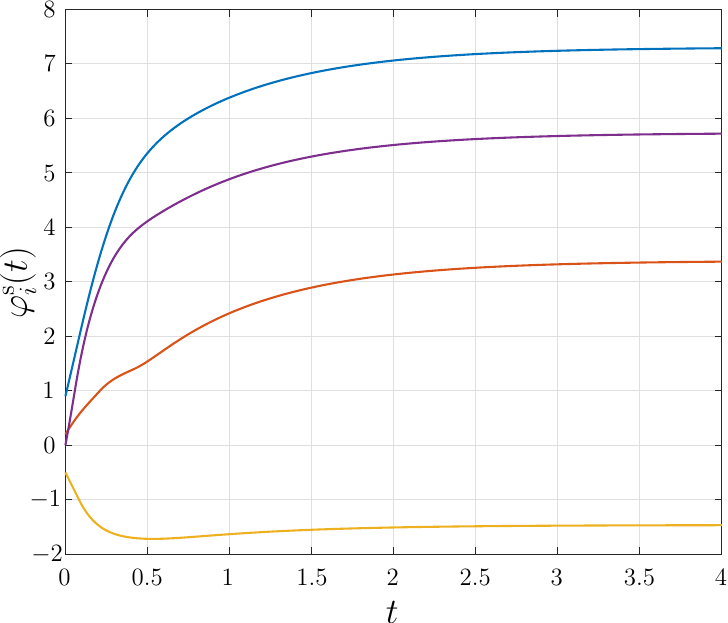}
  \caption{\small }
\end{subfigure} \\
\begin{subfigure}{0.47\columnwidth}
    % include first image
  \includegraphics[width=\linewidth]{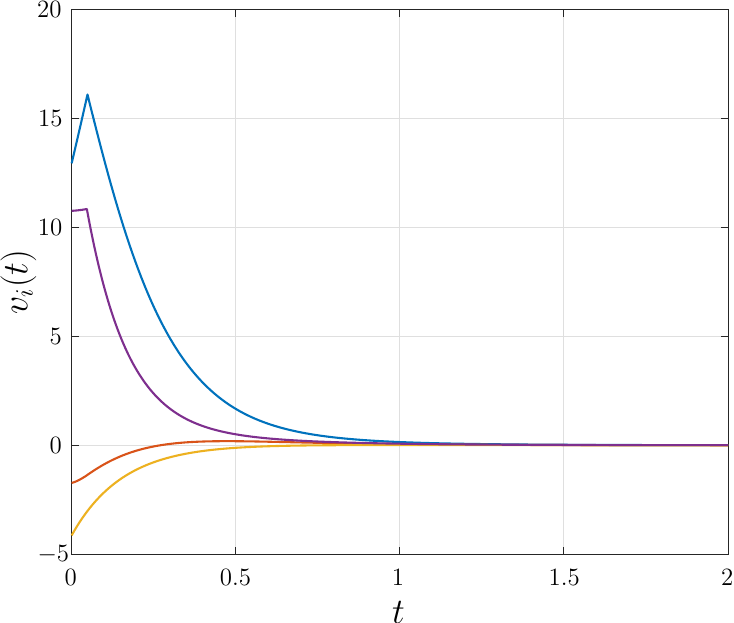}
  \caption{\small }
\end{subfigure}
\begin{subfigure}{0.47\columnwidth}
    % include second image
  \includegraphics[width=\linewidth]{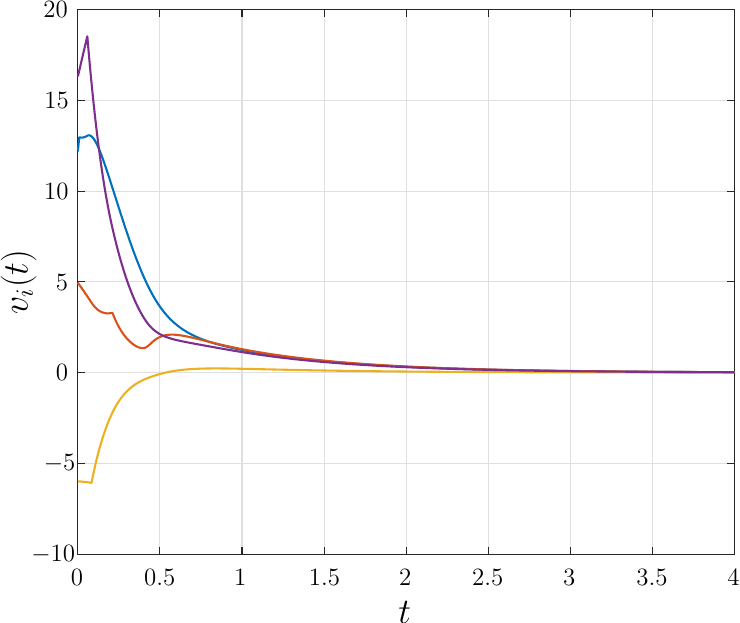}
  \caption{\small }
\end{subfigure}\\
\begin{subfigure}{0.47\columnwidth}
  % include first image
  \includegraphics[width=\linewidth]{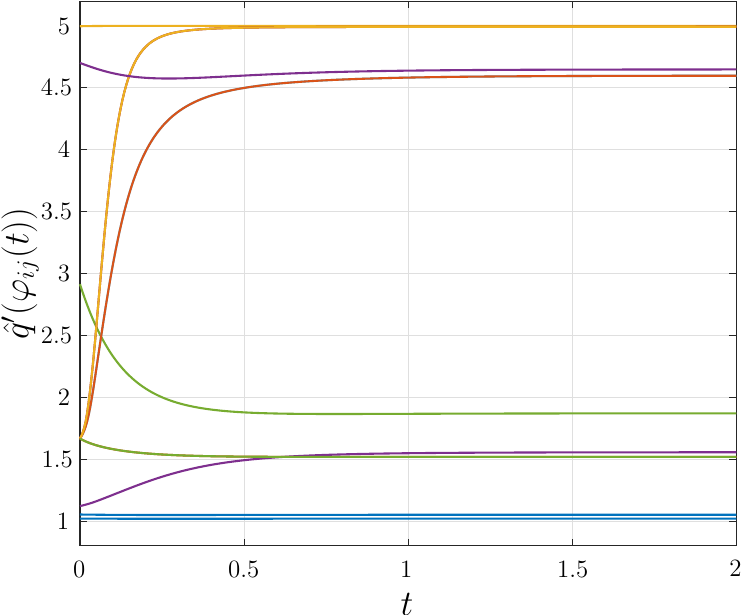}
  \caption{\small }
\end{subfigure}
\begin{subfigure}{0.47\columnwidth}
    % include second image
  \includegraphics[width=\linewidth]{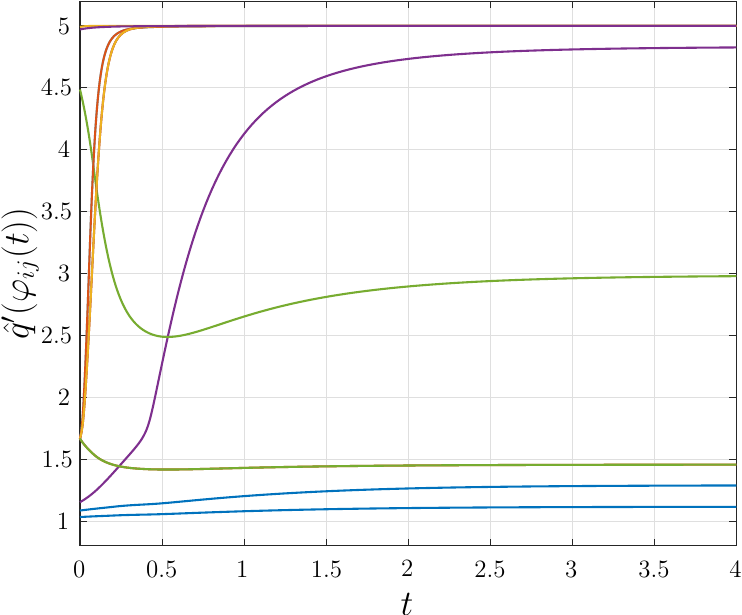}
  \caption{\small }
\end{subfigure} \\
\caption{\small (a), (b) Time-domain behavior of fluxes $\varphi_i^\mathrm{s}$ for a fourth-order MNN
with three-segments pwl memristor self-connections and interconnections implemented via an HP memristor. Two different sets of
initial conditions $\varphi_{i0}^\mathrm{s}$, $i=1,\dots,4$, are considered ($\alpha=8$). (c), (d) Corresponding behavior of activations $v_i$ and (e), (f) of memductances $\hat q'(\varphi_{ij})$.}
\label{fig:trans beta1 HP}
\end{center}
\end{figure}

Figures\ \ref{fig:trans beta1 HP}(a), (b) show the time-domain
behavior of the neuron fluxes $\varphi_i^\mathrm{s}$ when $\alpha =8$ for
two different sets of initial conditions $\varphi_{i0}^\mathrm{s}$, $i=1,\dots,4$.
Note that we have convergence to two different EPs.
Figures\ \ref{fig:trans beta1 HP}(c), (d) show the behavior of the neuron activations $v_i$ of (\ref{MNN}), while Figs.\ \ref{fig:trans beta1 HP}(e), (f) show the behavior of the interconnection memductances
$\hat q'(\varphi_{ij})$. In this case the memductances, i.e., the interconnection weights, are
subject to smooth changes before convergence.

\subsection{2D MNN for a Simple Image processing Task}
\label{sect:MNN image proc}

We consider a 2D MNN to solve the task of hole filling for a $20 \times 20$
image. Now, each neuron is connected
only with the two adjacent neurons in the same row and the two adjacent ones in the
same column. With this choice, the interconnection structure is akin to that of a standard
cellular NN \cite{kek2007cellular} for solving an analogous task. Again,
let us  put the MNN equations in the form (\ref{MNN}) in the
VCD and in the form (\ref{MNNFC}) in the FCD. The indicator matrix $T$
is irreducible. Let us choose $Q_0=0$, $C_i=1$ and $G_i+G_{ai}=G=-3$, $i=1,\dots,20$.
Also in this case the memristors have a
three-segment pwl characteristic. For
the memristor self-connections we let $G_-^\mathrm{s}=G_+^\mathrm{s}=10$,
$G_0^\mathrm{s}=0.1$, while for the memristor interconnections we have
$G_-=G_+=0.69$, $G_0=0.345$. For all memristors, $\sigma_-=-1$ and $\sigma_+=1$.
It can be checked that with these choices the condition
for bounded solutions is satisfied.
The image processing is performed
in the FCD by the MNN (\ref{MNNFC}).
Now, we let $\varphi_{i0}^\mathrm{s}=-1.2$ for all $i$, which corresponds to a black
initial image.
The image to be processed is provided via the manifold index $Q_0$. Namely,
we choose $Q_{i0}=3.8 \times 1.2$ (resp., $Q_{i0}=-3.8 \times 1.2$) if the
$i$-th pixel of the input image is white (resp., black).

\begin{figure}
\begin{center}
\begin{subfigure}{0.47\columnwidth}
  % include first image
  \includegraphics[width=\linewidth]{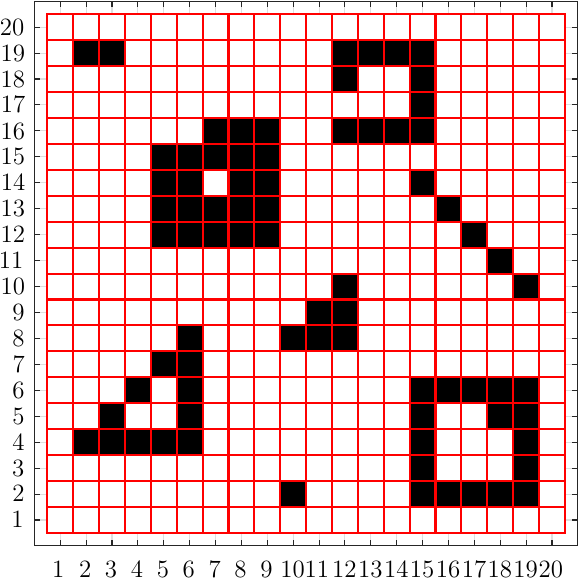}
  \caption{\small }
\end{subfigure}
\begin{subfigure}{0.47\columnwidth}
    % include first image
  \includegraphics[width=\linewidth]{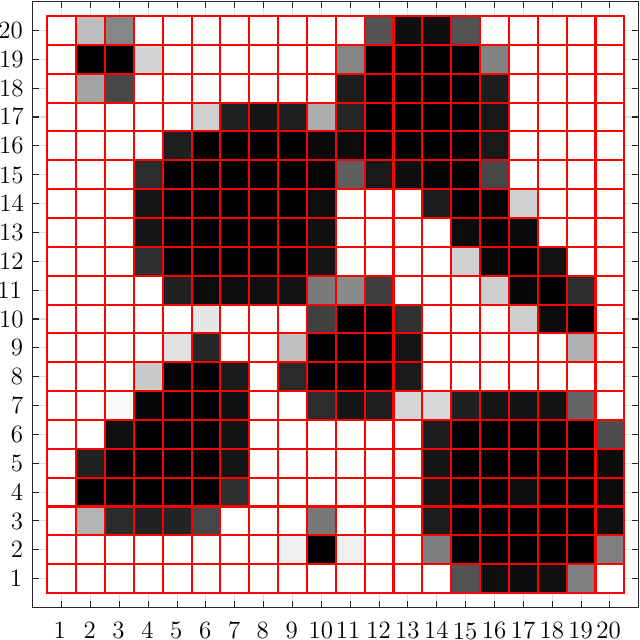}
  \caption{\small }
\end{subfigure}\\
\begin{subfigure}{0.47\columnwidth}
    % include second image
  \includegraphics[width=\linewidth]{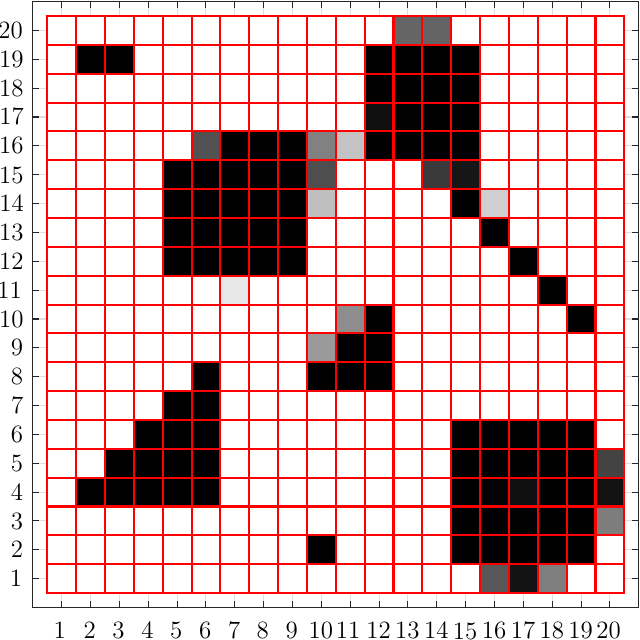}
  \caption{\small }
\end{subfigure}
\begin{subfigure}{0.47\columnwidth}
    % include second image
  \includegraphics[width=\linewidth]{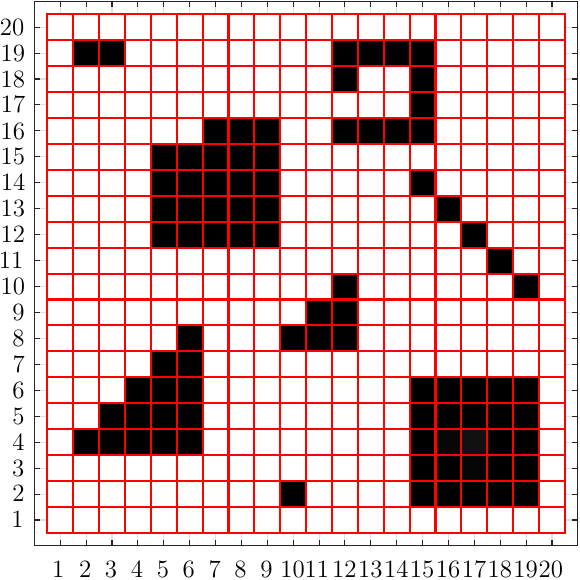}
  \caption{\small }
\end{subfigure} \\
\caption{\small \small 2D MNN for solving an hole filling task. a) Initial image to be processed, b) and
c) intermediate snapshots at $t=1.4$, $t=2.2$ and d) final image after processing.}
\label{fig:MNN HF}
\end{center}
\end{figure}

\begin{figure}
\centering
\begin{subfigure}{0.49\columnwidth}
  % include first image
  \includegraphics[width=\linewidth]{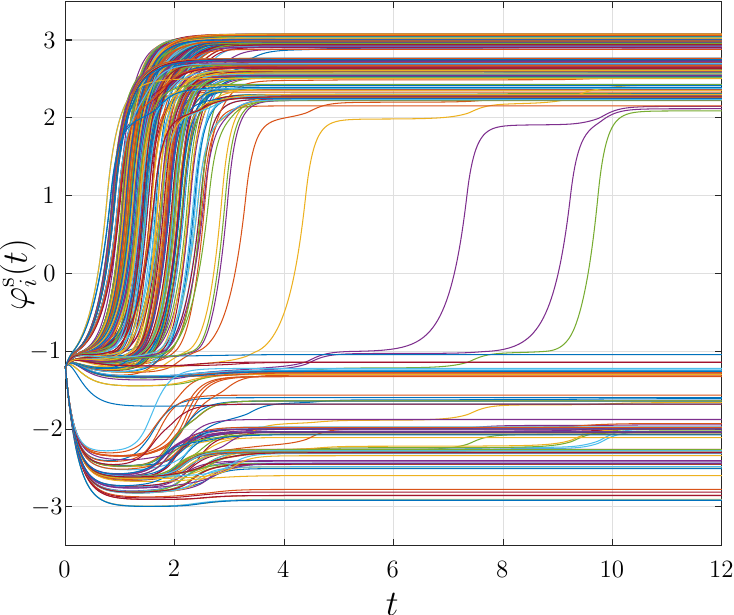}
  \caption{\small }
\end{subfigure}
\begin{subfigure}{0.49\columnwidth}
    % include second image
  \includegraphics[width=\linewidth]{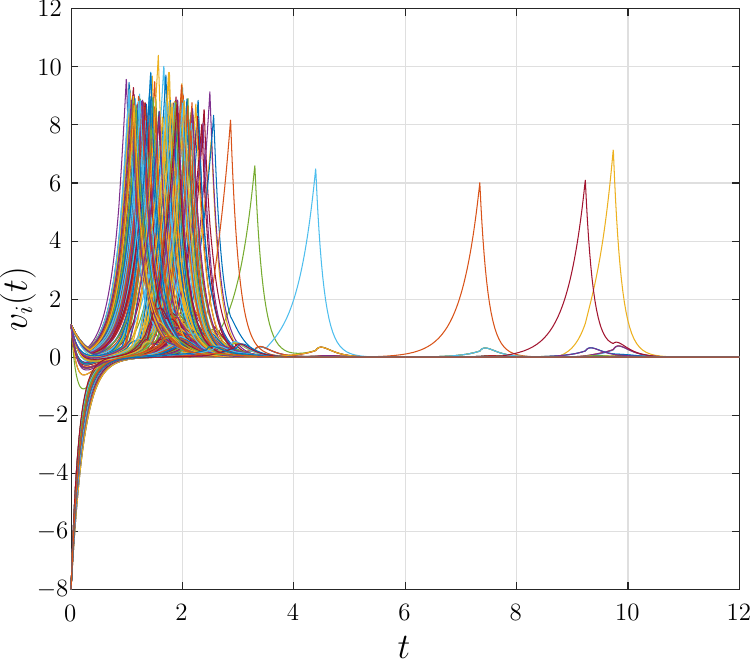}
  \caption{\small }
\end{subfigure} \\
\begin{subfigure}{0.49\columnwidth}
    % include first image
  \includegraphics[width=\linewidth]{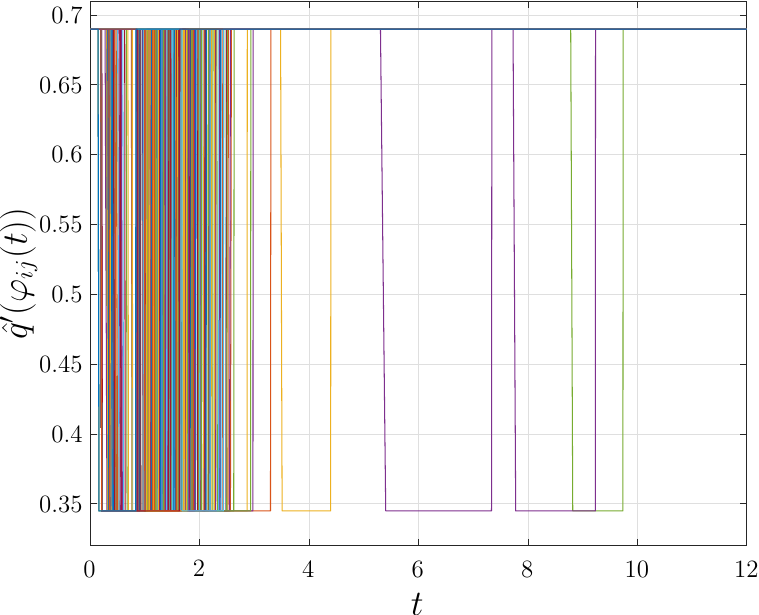}
  \caption{\small }
\end{subfigure}
\caption{\small (a) Time-domain behavior of fluxes $\varphi_i^\mathrm{s}$ for a 2D MNN
for solving a hole filling task, (b) behavior of neuron activations $v_i$ and
(c) behavior of interconnection menductances $\hat q'_{ij}(\varphi_{ij})$.}
\label{fig:trans HF}
\end{figure}

Figure\ \ref{fig:MNN HF} shows the input image, the final image and two intermediate snapshots during the transient. It is seen that the MNN correctly solves the hole filling task. The evolution of the fluxes $\varphi_i^\mathrm{s}$, activations $v_i$ and memductances weights $\hat q'(\varphi_{ij})$ for the MNN (\ref{MNN}) in the VCD is shown in Fig.\ \ref{fig:trans HF}. Note once more that $v_i \to 0$ and that we can observe several memductance switchings before the MNN settles.

\section{Discussion}
\label{sect:disc}

\begin{remark}
We have proved convergence of the MNN (\ref{MNNFC}) in the FCD by means of the theory
of cooperative systems of differential equations by Hirsch \cite{Hir85,Hir89}.
It is worth to stress that, as it can be easily verified, the MNN (\ref{MNN}) is not cooperative in the standard VCD, hence its convergence properties cannot be
directly derived from the theory of cooperative systems. This shows that the use of FCAM to
obtain a cooperative reduced system (\ref{MNNFC}) in the FCD on each invariant manifold
is crucial for the approach to convergence in the paper.
\end{remark}

\begin{remark}
One main assumption to apply the results on convergence in this paper is that there are
positive feedback interconnections between distinct neurons (cf.\ Sect.\ \ref{sect:mem model}).
It is pointed out that there are other classes of MNNs that can be brought back to this case.
A relevant one is that of MNNs that are sign-symmetric and consistent according to the
definition in \cite[Sect.\ 11]{Hir89}. In that case, a suitable change of variables permits
to transform such MNNs in a MNN with positive neuron interconnections. We refer the reader to
\cite{Hir89,SW02} for more details.
\end{remark}

\begin{remark}
The main result on convergence here obtained is a broad generalization of previous results on convergence for memristor MNNs obtained in \cite{7516733}. The main difference is in that paper memristors are used to model the neurons, while the neuron interconnections are implemented via resistors and as such they are constant. Therefore, \cite{7516733} deals with convergence of the activation dynamics, while in this paper we are dealing with the general case of time-varying memristor interconnections and convergence of the weight-activation dynamics.
\end{remark}

\begin{remark}

A number of papers in the literature have considered NNs where one or more
memristors are used for the interconnections, see, e.g., \cite{Ding2023,Bao202310526,Chen20242308,Lai20244039} and references therein. Some main dynamic properties there studied
concern complex dynamics, chaos and synchronization phenomena implied by the presence of
dynamic memristors. As an example, the article \cite{10884532} investigates on the use of two memristor
interconnections in a unidirectional ring of Hopfield-type NNs enabling to enhance the
complex NN behavior. The article \cite{9709550} considers instead a low-order Hopfield-type NN
where a memristor interconnection is used for generating multi-scroll chaotic attractors.
\end{remark}

\begin{remark}
It is worth to mention that there is a branch of the literature concerning the dynamic
analysis of NNs where the
neuron interconnections are implemented via a basically different memristor model with respect to
the standard ideal memristor model by Chua \cite{Chua1971} considered in the MNN (\ref{MNN}).
See, e.g., \cite{yang2016exponential, 6877732,LIU20204437,Deng20225367,Zhang20237844} and their references.
More precisely, those papers
use a memristor modeled as a nonlinear resistor that switches between two different
values of the memductance. As it is discussed in \cite{pershin2020validity}, it is not clear however how that model relates to the ideal memristor model by Chua.
\end{remark}

\section{Conclusion}

The paper has considered a class of NNs with memristor interconnections (MNNs) where both the neuron
activations and interconnections are allowed to vary with time (weight-activation dynamics).
A general result on convergence has been proved in the case of positive feedback memristor
interconnection under an irreducibility assumption on the interconnections. The result includes
the most interesting case in view of the applications where the MNNs have multiple
stable EPs and it is robust with respect to perturbations.
To the authors' knowledge, this is the first general convergence result
for weight-activation dynamics. The obtained result has been verified and illustrated via the
application to some simple processing tasks in real time.
This is only a first step in the investigation of what we feel is a new and broad topic concerning convergent weight-activation dynamics. Future work
will be devoted to study if it is possible to obtain convergence results for the
weight-activation dynamics when dropping the assumption of positive memristor interconnections
or when using different interconnection structures for the NNs.

\section*{Appendix A}
\label{app:bound}

Function $V(\varphi^\mathrm{s})=\|\varphi^\mathrm{s}\|_\infty:\R^n \to \R$ is convex and Lipschitz with Lipschitz constant 1 in $\R^n$. Then, by the chain rule as in \cite[Property\ 1]{FGNP06} we have
\begin{align}
\label{eq:scalarprod}
\dot V(\varphi^\mathrm{s}(t)) & =\langle \xi, \dot{\varphi}^\mathrm{s}(t) \rangle, \ \ \forall \xi \in \partial V (\varphi^\mathrm{s}(t))
\end{align}
for almost all $t\in [0,\tau]$, where $\langle \cdot, \cdot \rangle$ denotes the scalar product, and $\partial V(\varphi^\mathrm{s}(t))$ is the subgradient in the sense of convex analysis of $V$ evaluated at $\varphi^\mathrm{s}(t)$ \cite[Def.\ 1.2.1]{convex}. Recall that the index $\bar{k}\in \mathfrak{N}$ is such that $|\varphi^\mathrm{s}_{\bar k}(t)|=\|\varphi^\mathrm{s}(t)\|_\infty=\max_{k \in \mathfrak{N}}|\varphi^\mathrm{s}_k(t)|$. Then, for any $\varphi^\mathrm{s}_k(t)\ne 0$, the vector $\xi(t)$ defined as $\xi_{\bar k}(t)= \mathrm{sgn}(\varphi^\mathrm{s}_{\bar k}(t))$, $\xi_i(t)=0$, 
$i\in \mathfrak{N}, i \neq \bar{k}$, is such that $\xi_{\bar k}(t) \in \partial V (\varphi^\mathrm{s}(t))$.
From~(\ref{MNNFC}), it is possible to write~(\ref{eq:scalarprod}) as
$
\dot V(\varphi^\mathrm{s}(t)) = -(G_{a \bar{k}}+G_{\bar k})\varphi^\mathrm{s}_{\bar k}(t) \mathrm{sgn}(\varphi^\mathrm{s}_{\bar k}(t)) - \hat{q}_{\bar k}(\varphi^\mathrm{s}_{\bar k}(t)) \mathrm{sgn}(\varphi^\mathrm{s}_{\bar k}(t))
+ \sum_{j \in C_{\bar k}}\hat{q}_{\bar{k}j}(\varphi^\mathrm{s}_{j}(t)-\varphi^\mathrm{s}_{j0}+\varphi_{\bar{k}j0}) \mathrm{sgn}(\varphi^\mathrm{s}_{\bar k}(t)) + Q_{\bar{k}0} \mathrm{sgn}(\varphi^\mathrm{s}_{\bar k}(t)).
$
Since $\hat{q}(0)=0$ and $\hat{q}'>0$ (cf.\ Assumption~\ref{assu:mem}), we have that $\mathrm{sgn}(\varphi^\mathrm{s}_{\bar k}(t))=\mathrm{sgn}(\hat{q}_{\bar k}(\varphi^\mathrm{s}_{\bar k}(t)))$, hence we have
\begin{equation}
\label{eq:majder}
\begin{split}
\dot V(\varphi^\mathrm{s}(t)) \le & -(G_{a \bar{k}}+G_{\bar k})|\varphi^\mathrm{s}_{\bar k}(t)| - |\hat{q}_{\bar k}(\varphi^\mathrm{s}_{\bar k}(t))| \\
 &+ \sum_{j \in C_{\bar k}}|\hat{q}_{\bar{k}j}(\varphi^\mathrm{s}_{j}(t)-\varphi^\mathrm{s}_{j 0}+\varphi_{\bar{k}j0})| + |Q_{\bar{k}0}| \\
\le & -(G_{a \bar{k}}+G_{\bar k})|\varphi^\mathrm{s}_{\bar k}(t)| \\
 &- \min \{|\hat{q}_{\bar k}(\varphi^\mathrm{s}_{\bar k}(t))| ,|\hat{q}_{\bar k}(-\varphi^\mathrm{s}_{\bar k}(t))| \} \\
 &+ \sum_{j \in C_{\bar k}} \max \{|\hat{q}_{\bar{k}j}(\varphi^\mathrm{s}_{j}(t)-\varphi^\mathrm{s}_{j0}+\varphi_{\bar{k}j0})|,\\
 & |\hat{q}_{\bar{k}j}(-\varphi^\mathrm{s}_{j}(t)-\varphi^\mathrm{s}_{j0}+\varphi_{\bar{k}j0})| \}
+ |Q_{\bar{k}0}|.
\end{split}
\end{equation}
Function $\hat{q}_{\bar{k}j}$ is strictly monotone increasing and $|\varphi^\mathrm{s}_{\bar k}(t)|=\|\varphi^\mathrm{s}(t)\|_\infty \ge \varphi^\mathrm{s}_{j}(t)$. Hence
$
\max \{|\hat{q}_{\bar{k}j}(\varphi^\mathrm{s}_{j}(t)-\varphi^\mathrm{s}_{j0}+\varphi_{\bar{k}j0})| ,|\hat{q}_{\bar{k}j}(-\varphi^\mathrm{s}_{j}(t)-\varphi^\mathrm{s}_{j0}+\varphi_{\bar{k}j0})| \}
\le \max \{|\hat{q}_{\bar{k}j}(\varphi^\mathrm{s}_{\bar{k}}(t)-\varphi^\mathrm{s}_{j0}+\varphi_{\bar{k}j0})| ,|\hat{q}_{\bar{k}j}(-\varphi^\mathrm{s}_{\bar{k}}(t)-\varphi^\mathrm{s}_{j0}+\varphi_{\bar{k}j0})|\}.
$
Using this inequality in~(\ref{eq:majder}) leads to~(\ref{DVDT}). \qed

\subsection*{Biographies}

{\bf M. Di Marco}
was born in Firenze, Italy, in 1970. He received the
Laurea Degree in Electronic Engineering from the University of Firenze,
Firenze, Italy and the Ph.D. degree from the University of Bologna, Bologna,
Italy, in 1997 and 2001, respectively. From November 1999 to
April 2000 he held a position as Visiting Researcher at LAAS, Toulouse,
France. Since 2000, Dr. Di Marco has been with the University of Siena, Siena,
Italy, where he is currently Associate Professor of Circuit Theory. He is co-author of more than 110 international journal and conference papers. His current research interests are in analysis and modeling of nonlinear dynamics of complex systems and neural networks, in robust estimation and filtering. From 2007 to 2011, Dr. Di Marco has been serving as Associate Editor of the IEEE TRANSACTIONS ON CIRCUITS AND SYSTEMS I: REGULAR PAPERS. From 2020, he is serving as Associate Editor of the Springer Neural Processing Letters.
\vskip.2cm
{\bf M. Forti}
received the Laurea degree in Electronics Engineering
from the University of Florence, Italy, in 1988. From
1991 to 1998, he was an Assistant Professor in applied mathematics
and network theory with the Electronic Engineering Department of
the University of Florence. In 1998 he joined the Department of
Information Engineering and Mathematics
of the University of Siena, Italy, where
he is currently Professor of electrical engineering.
His main research interests are in the field of nonlinear circuits
and systems, with emphasis on the qualitative analysis and
stability of circuits modeling artificial neural networks. His
research activity also includes aspects of electromagnetic
compatibility. Dr. Forti served as Associate Editor of the
IEEE TRANSACTIONS ON CIRCUITS AND SYSTEMS I: FUNDAMENTAL THEORY
AND APPLICATIONS from 2001 to 2003 and of the IEEE TRANSACTIONS
ON NEURAL NETWORKS from 2001 to 2010. Since 2016 he is serving as
Associate Editor of the IEEE TRANSACTIONS ON CYBERNETICS. Since
2024 he is also serving as a Co-Editor in Chief of
Neural Networks (Elsevier).
\vskip.2cm
{\bf L. Pancioni}
received the Laurea degree in telecommunication engineering
from the University of Siena, Siena, Italy, in 2001 and the Ph.D. degree in
information engineering from the University of Siena, Siena, Italy in 2004.
He is currently Assistant Professor of Electrical Engineering at the
Department of Information Engineering and Mathematics of Siena University.
His main research interests are in the field of analysis of nonlinear circuits
modeling neural networks, mainly focused on stability and complex dynamics.
Recent activities concern the study and the modeling of neural
circuits including memristors. His research activity also includes modeling
of source-coupled logic and electronic design of integrated analog and
mixed signals.
\vskip.2cm
{\bf G. Innocenti}
graduated in 2004 with a Master's Degree in Information
Engineering at the University of Florence (Firenze, Italy), and from the
same institution received the Ph.D. in Nonlinear Dynamics and Complex
Systems in 2008. He was a Postdoctoral Research Fellow, first at the
University of Florence and then at the University of Siena (Siena, Italy),
from 2008 to 2010. Since 2012 he has been Assistant Professor in Automation
at the University of Florence, where he is with the Department of
Information Engineering. He serves as Associate Editor for scientific
journals in the field of the analysis of nonlinear systems, and he was in
the committees of workshops and scientific congresses on the same subject.
His main research is nonlinear dynamics with particular interest in
networks of interacting agents and neuron models.
\vskip.2cm
{\bf A. Tesi}
received the Laurea degree in Electronics Engineering from the
University of Florence, in 1984, and the Ph.D. degree in Systems Engineering
from the University of Bologna, in 1989. In 1990 he joined the Department of
Systems and Computer Science of University of Florence, as a research assistant.
He is currently a Professor of Control Systems at the Department of Information
Engineering of University of Florence. He was associate editor of the
IEEE TRANSACTIONS ON CIRCUITS AND SYSTEMS (1994-1995), IEEE TRANSACTIONS ON
AUTOMATIC CONTROL (1995-1998) and Systems and Control Letters (1995-2010). He
was member of the conference editorial board of the Conference on
Decision and Control (1994-1999) and the American Control Conference (1995-2000),
and member of the program committee of several international conferences.
His research interests are in analysis of nonlinear dynamics of complex
systems, robust control of linear systems and optimization. He is co-author
of about 180 scientific publications.

\end{document}